\documentclass[aps,prb,superscriptaddress,preprint]{revtex4-1}

\usepackage{graphicx}
\usepackage{subfigure}
\usepackage{caption}
\usepackage{epstopdf}
\usepackage{epsfig}

\usepackage{amssymb}
\usepackage{amsthm}

\begin{document}


\title{DC Magnetization study of Ni$_{1-x}$Rh$_x$ nanoalloy}


\author{P. Swain}
\email[]{priya@phy.iitkgp.ernet.in}
\affiliation{Department of
Physics, Indian Institute of Technology Kharagpur,
Kharagpur-721302, INDIA}
\author{A. Thamizhavel}
\affiliation{Dept. of Condensed Matter Physics \& Materials
Science, Tata Institute of Fundamental Research, Colaba, Mumbai
400005, INDIA }

\author{Sanjeev K. Srivastava}
\email[]{sanjeev@phy.iitkgp.ernet.in} \affiliation{Department of
Physics, Indian Institute of Technology Kharagpur,
Kharagpur-721302, INDIA}



\begin{abstract}
 Ni$_{1-x}$Rh$_x$ bulk alloys  exhibit a ferromagnetic to paramagnetic  quantum phase transition (QPT) at the critical concentration
$x_c \sim$ 0.37. A spin glass phase arises below the
Ferromagnetism by controlling the short range interaction between
Ni and Rh atoms. we have synthesized nanoalloys of two different
concentrations with two different reaction times by chemical
reflux method. From the dc magnetization measurements, existence
of ferromagnetism and spin glass phase in these nanoalloys is
confirmed. A step like feature  in  both ZFC and FC curve at
certain temperature indicates the existence of the spin glass
phase.

\end{abstract}

%
%
%

%
%

\maketitle

\section{Introduction}
A quantum phase transition (QPT), a continuous order disorder
transition at zero temperature, is driven by quantum fluctuations.
A ferromagnetic to paramagnetic QPT in the Ni can be archived by
adding different non magnetic impurities like Pd,\cite{Nicklas99}
V, \cite{Schroeder10} Pt \cite{Ododo77} and Rh \cite{Muellner}.
Out of all these systems the Ni$_{1-x}$Rh$_x$ is the system in
which the magnetic phase transition is quite complicated. The
Ni$_{1-x}$Rh$_x$ system exhibit a QPT from ferromagnetic to
paramagnetic  below a critical composition of 0.626,
\cite{Carnegie93, Muellner, Krishnamurthy94} The onset of
ferromagnetism in these systems are spin glass ordering. In
Ni$_{1-x}$Rh$_x$ below the percolation threshold of the
ferromagnetism a spin glass phase appears due to the interaction
between the polarizing spin clusters which form at the Ni rich
region.\cite{Carnegie93}Bulk Ni$_{1-x}$Rh$_x$ alloy show short
range ordering which means Ni atom is preferred  to be surrounded
by Rh atoms however in case of homogenous Ni$_{1-x}$Rh$_x$ alloy,
the system becomes more random. So the no of Ni nearest neighbor
to a Ni given atom increases, giving rise to an increase in Ni
clusters. The interaction between such clusters move forward the
system from paramagnetic to spin glass type below critical
concentration of ferromagnetic order. The heat capacity and the
magnetic susceptibility show anomalous behavior at the critical
concentration in low temperature region.
 Materials at nanoscale are known to possess properties different from their bulk counterparts.
 Specifically studying the magnetic QPT of the above well studied bulk alloy in the nanoscale is very much complicated since the
 nanosystem itself inherited by multitude of magnetic phases, viz. paramagnetic, superparamagnetic, blocked
 ferromagnetic. So It is quite enthralling,to investigate the  QPT in Ni$_{1-x}$Rh$_x$ nanoallys.
  The synthesis and the catalytic property of Ni$_{1-x}$Rh$_x$
 nanoalloys  were studied previously. Recently lots of studies on Ni$_{1-x}$Rh$_x$  nanoalloys and graphene supported nanoalloys are going on due to its high use
  as catalyst in hydrogen storage \cite{Bingquan, Junfeng, Wang} and hydrogen generation. \cite{Changming, Pingping, Zhujun}
  However any type  report addressing to the  magnetic  study or possibility of composition driven magnetic QPT in Ni$_{1-x}$Rh$_x$ nanoalloys
 has not been discussed as the best of our knowledge. This leaves us an opportunity to study the magnetism and composition driven QPT in this nanoalloy.
 In this work, we have desired to study the magnetism of Ni$_{1-x}$Rh$_x$ nanoallys and the effects of the reaction time on the magnetism of the nanoalloys.
 We synthesize nanoparticles of Ni$_{1-x}$Rh$_x$  alloys above and below the bulk critical concentration, with keeping an aim of achieving good  crystallinity
 and single phased chemically with two different reaction times.
 We then  examined the size and crystallinity of the nanopaticles
 by microscopic and spectroscopic techniques. A further study on the magnetization on the nanoalloys was
 also carried out to get the idea about  the different magnetic properties of the nanoalloys.

\section{Experimental section}

\subsection{synthesis}

For synthesizing Ni, Rh, and  Ni$_{1-x}$Rh$_x$ alloy nanoparticles
a chemical reflux apparatus was employed using Rhodium(III)
chloride monohydrate (RhCl$_3$$.H_2$O), Nickel(II) chloride
(NiCl$_{2}$) as metal precursor and hydrazine hydrate as reducing
agent in the presence of surfactant diethanolamine. The method was
very similar to that reported by P.swain et. al \cite{swain15}
only some conditions are different. For a typical procedure, for
pure Rh or pure Ni 0.5 mmol of RhCl$_3$$.H_2$O or (NiCl$_{2}$) was
taken in a 100 mL round bottom flask containing 20 mL distill
water and dissolved completely. For Ni$_{1-x}$Rh$_x$ alloy
nanoparticles synthesis an appropriate ratio of both metal
precursors  were added subsequently. In this  step the $Ni^{+2}$
and $Rh^{+3}$ ions were generated. To the above mixed solution, 5
ml of diethanolamine was added as surfactant followed by 9 mL of
hydrazine hydrate as reducing agent.In the last reaction step, 30
ml distilled water was added to this.We have prepared two sets of
such solutions. The two sets were refluxed at 110 $^{\circ}$C one
for 6 hours and other set for 25 hours. Finally, the prepared
Ni$_{1-x}$Rh$_x$ alloy nanoparticles were separated by
centrifugation, washed with district water several times and dried
under vacuum for 48 hours. We denoted the prepared samples as N6
and N25 with 6 h, 25 h reaction time respectively. In this
nomenclature, 'N' stands for nanoparticles and the first numeric
is for reaction time in hours.In both the cases only refluxing
time was different whiles all other conditions were kept same.

\subsection{characterization}

Scanning electron microscope (SEM) by a Merlin  ZEISS scanning
electron microscope  and transmission electron microscope
measurement  by a JEOL JEM-2100 high resolution transmission
electron microscope operated at 200 kV  were carried out on the
Ni$_{1-x}$Rh$_x$ nanoalloy for the morphological and microscopical
study of the nanoalloys. For the HRTEM sample preparation,
nanoalloys were dispersed in acetone with 1 hour sonication. One
drop of the suspension solution was then placed on a piece of
carbon-coated copper grid. For SEM the pre sonicated solution was
dropped on small piece of aluminium sheet. Energy-dispersive X-ray
analysis (EDAX) was performed using a JEOL scanning electron
microscope for determining the compositions of synthesized alloys.
The X-ray diffraction (XRD) analysis were performed on a Philips
X-Pert MRD X-ray diffractometer with Cu K$_\alpha$ radiation to
confirm the structure and the phase.X-ray photoelectron
spectroscopy (XPS), using a PHI5000 Versaprobe system, was also
performed to further verify the stoichiometries of the samples.
Highly  Monochromatic focussed radiation from an Al-$K _\alpha$
source($h\nu$ = 1486.6 eV) X-ray source was used for excitation.
The pressure of the analyzer chamber was maintained in the range
of 1 $\times 10^{10}$ during the measurement. The binding energy
scale was charge referenced to C 1s at 284.5 eV. High-resolution
XPS spectra were acquired at 58.7 eV analyzer pass energy in steps
of 0.125 eV.

\subsection{Result and discussion}

\begin{table}
\caption{\label{tab1}Values of Rh concentration taken during synthesis $X_i$ and Rh concentration obtained from EDAX $X_s$  with the reaction time(t)in hours in the Ni$_{1-x}$Rh$_x$ nanoalloys.}
\centering
\begin{tabular}{lll}
\hline
$(t)   $&$X_i$&$X_s$\\
\hline
6&0.30&0.27 $\pm$0.02\\
6&0.65&0.64 $\pm$0.02\\
25&0.30& 0.269 $\pm$0.02\\
25& 0.65&0.64 $\pm$0.02\\
\hline
\end{tabular}\\

\end{table}

\subsection{EDAX}

Table 1 show the value of Rh concentration taken during the synthesis ($X_i$) and the Rh concentration obtained from the EDAX Quantification $X_s$ for the two different reaction times.
From the table we confirm that the values we got from EDAX is very close to the values taken during synthesis. The error $\sim$ 2$\%$ is taken as described by  Scott and Love.\cite{Scott1994}

\begin{figure}%
\centering
\subfigure[][]{%
\label{fig:1-a}%
\includegraphics[width=0.35\textwidth]{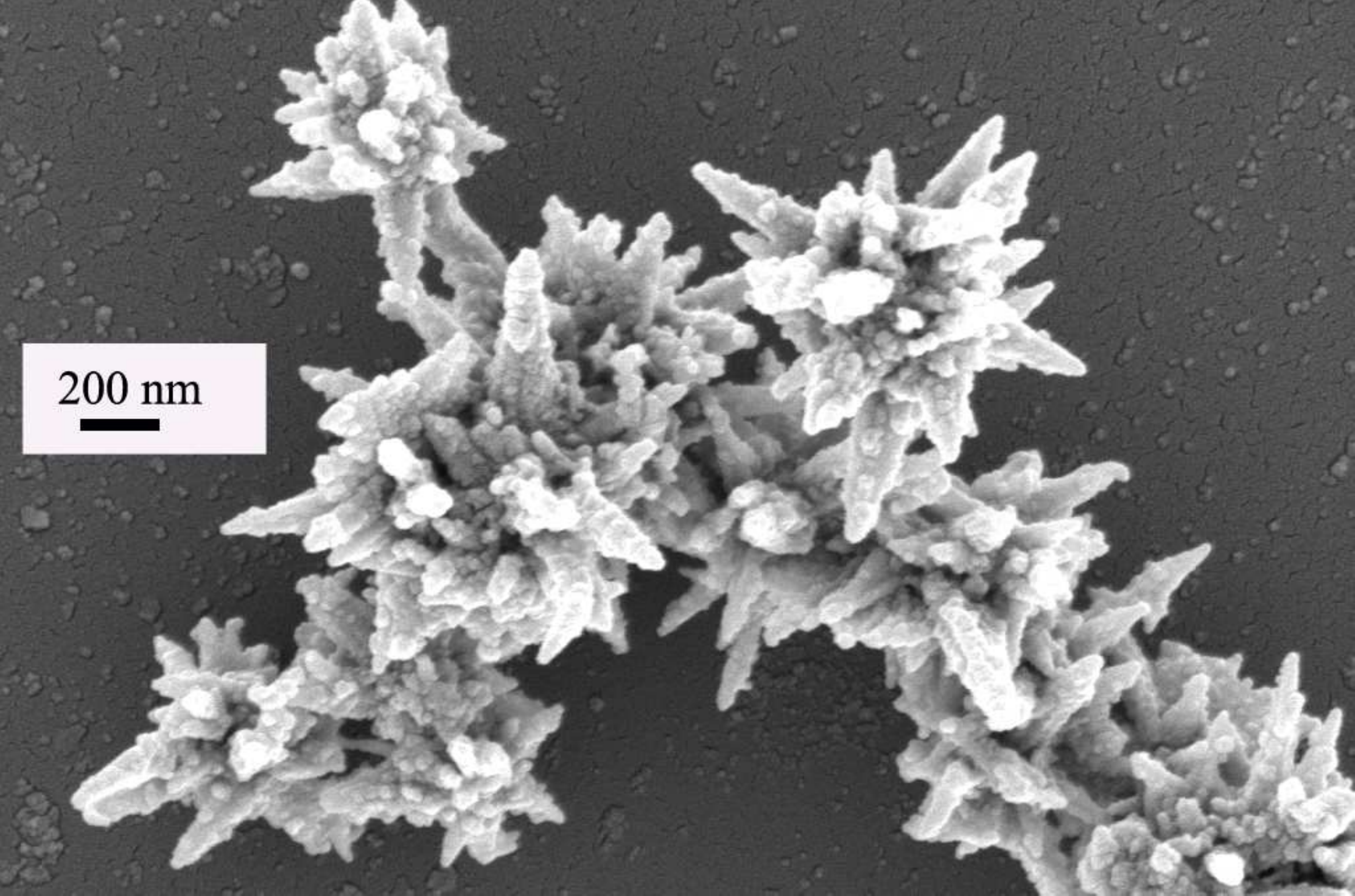}}%
\hspace{8pt}%
\subfigure[][]{%
\label{fig:1-b}%
\includegraphics[width=0.35\textwidth]{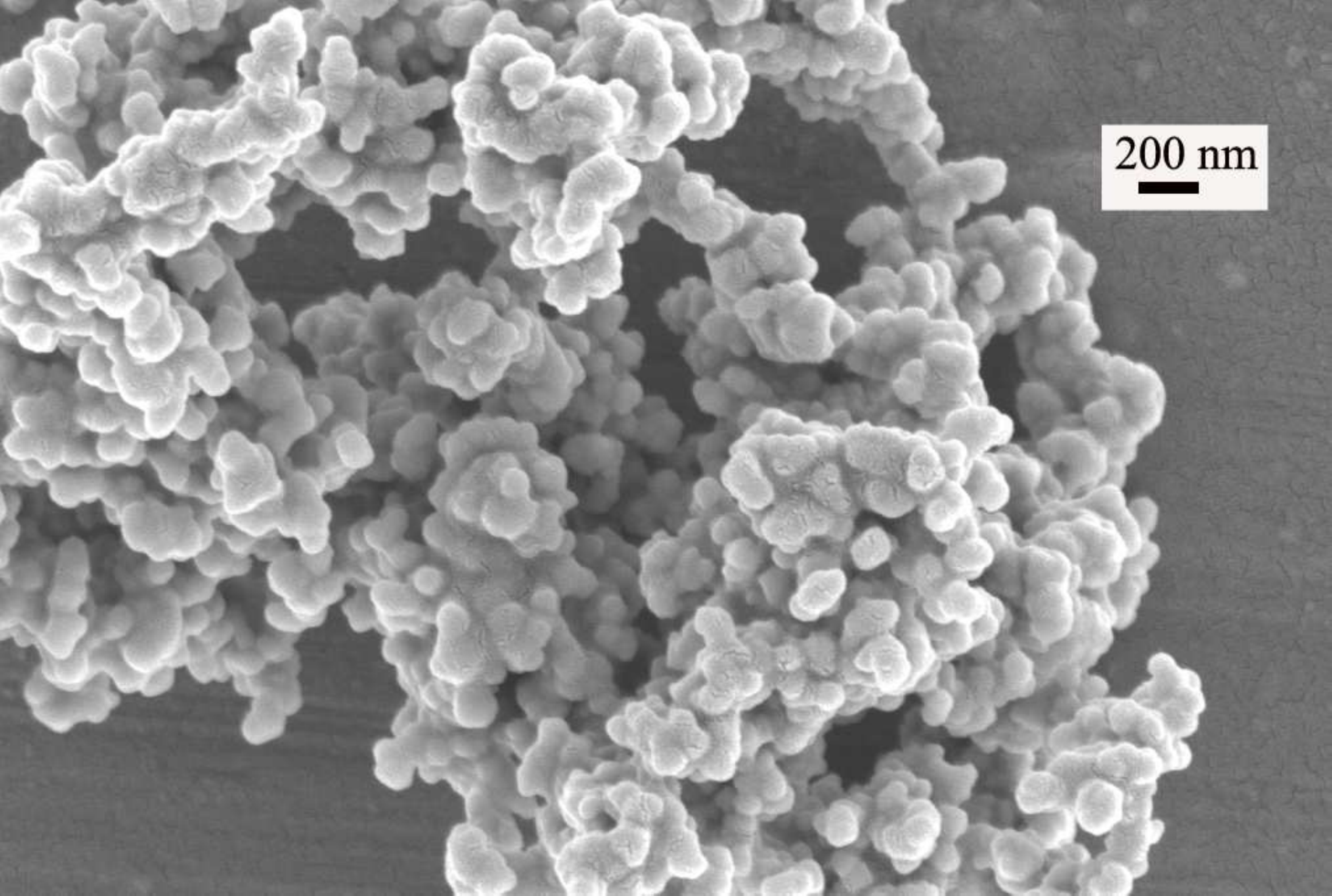}}%
\hspace{8pt}%
\subfigure[][]{%
\label{fig:1-c}%
\includegraphics[width=0.35\textwidth]{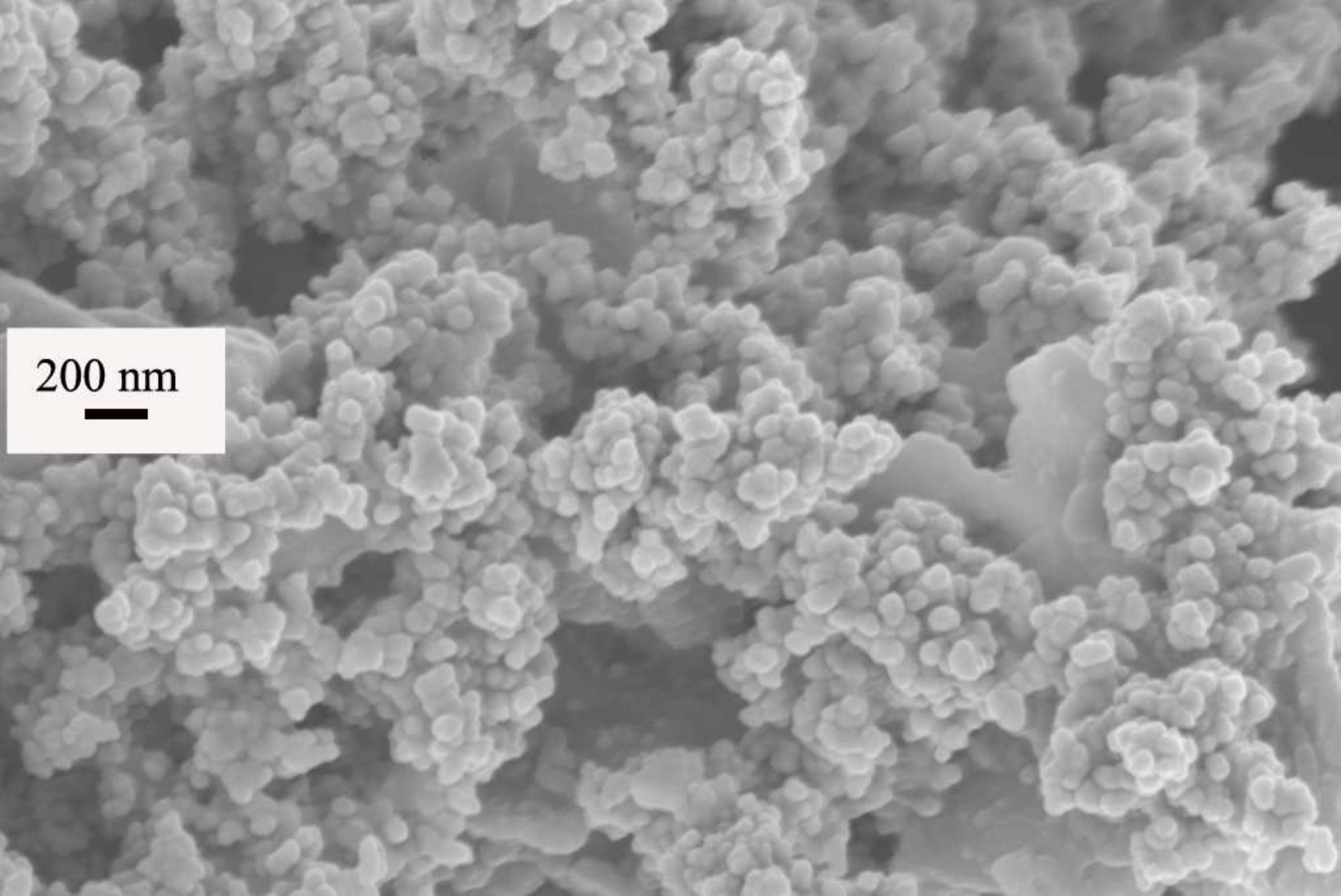}}%
\hspace{8pt}%
\subfigure[][]{%
\label{fig:1-d}%
\includegraphics[width=0.35\textwidth]{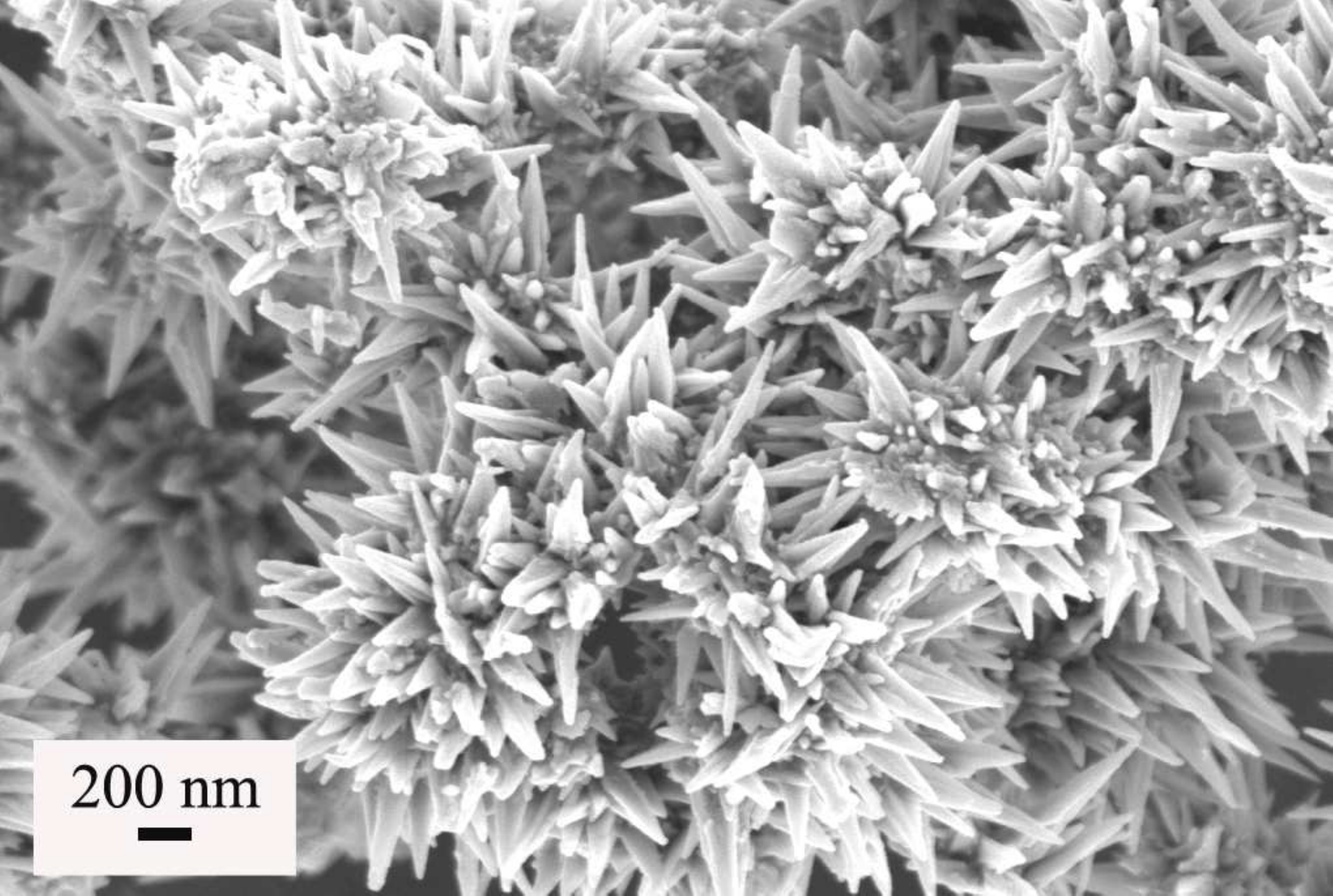}}%
\hspace{8pt}%
\subfigure[][]{%
\label{fig:1-e}%
\includegraphics[width=0.35\textwidth]{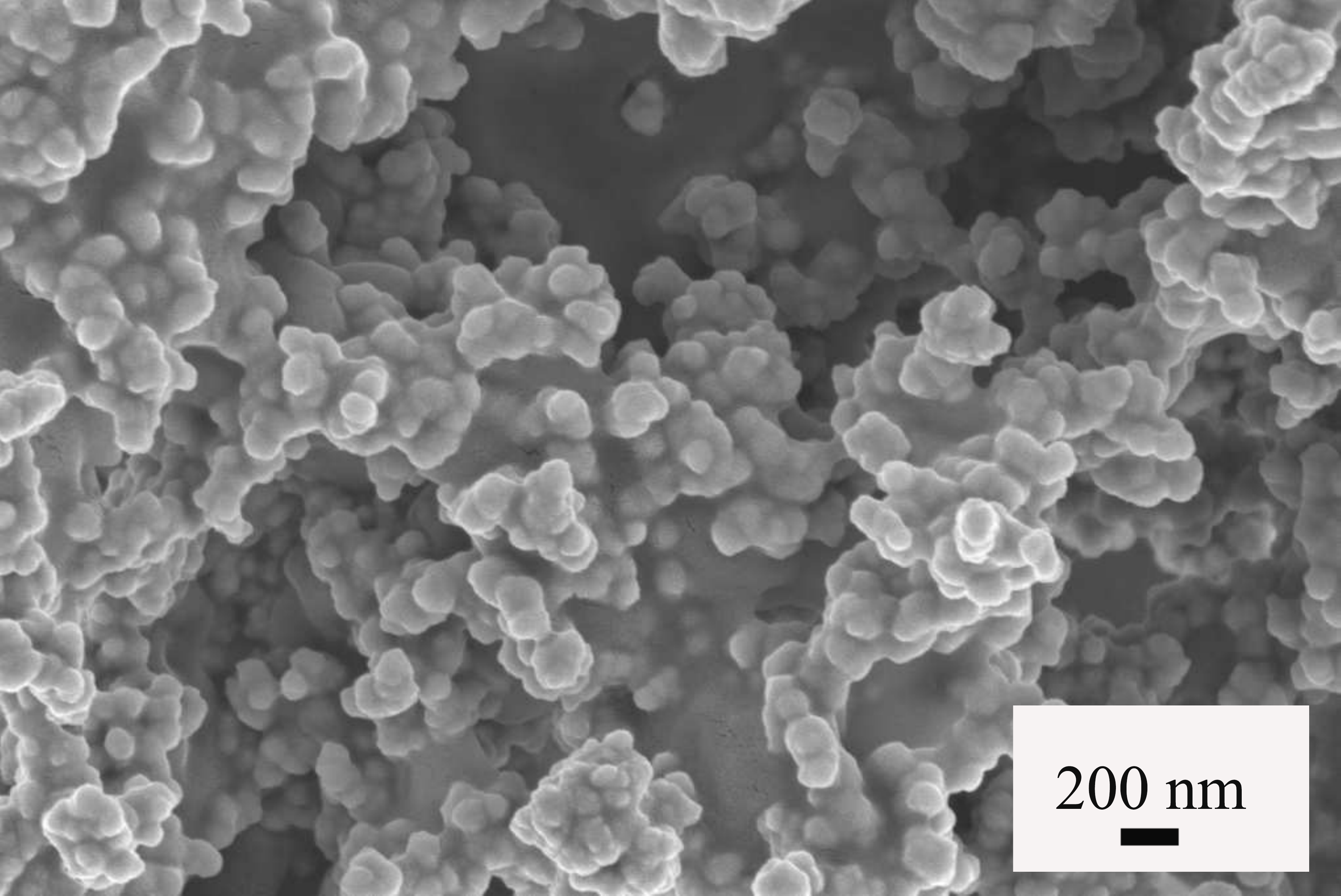}}%
\hspace{8pt}%
\subfigure[][]{%
\label{fig:1-f}%
\includegraphics[width=0.35\textwidth]{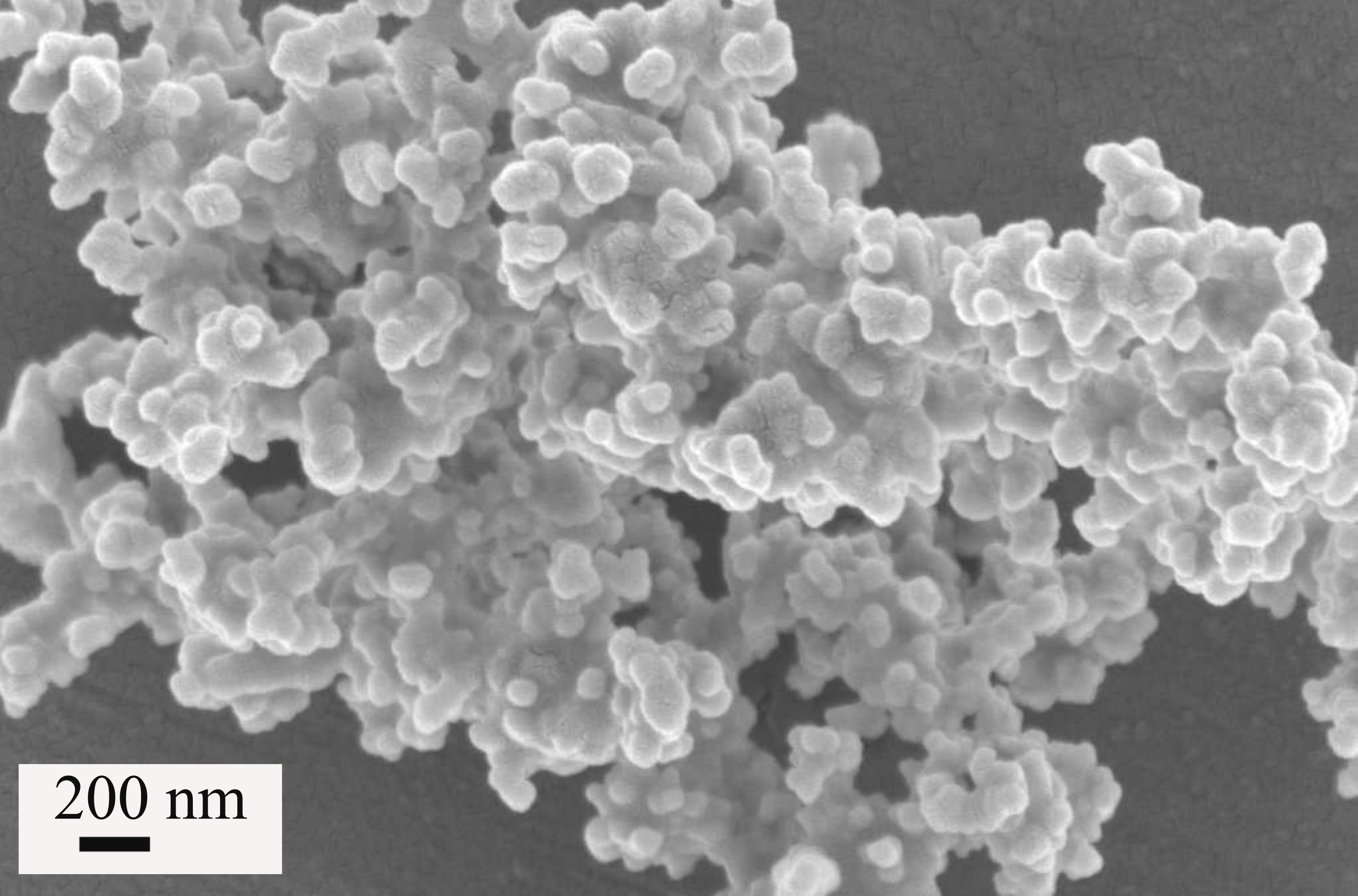}}%
\hspace{8pt}%

\caption[]{SEM images of Ni$_{1-x}$Rh$_x$ samples with Rh compositions $x$ = 0. (a), 0.27 (b), 0.64 (c) for N6, 0 (d) 0.27 (e) and 0.64 (f) for N25.}%
\label{fig:1}%
\end{figure}

\subsection{SEM and HRTEM}

\begin{figure}%
\centering
\subfigure[][]{%
\label{fig:2-a}%
\includegraphics[width=0.20\textwidth]{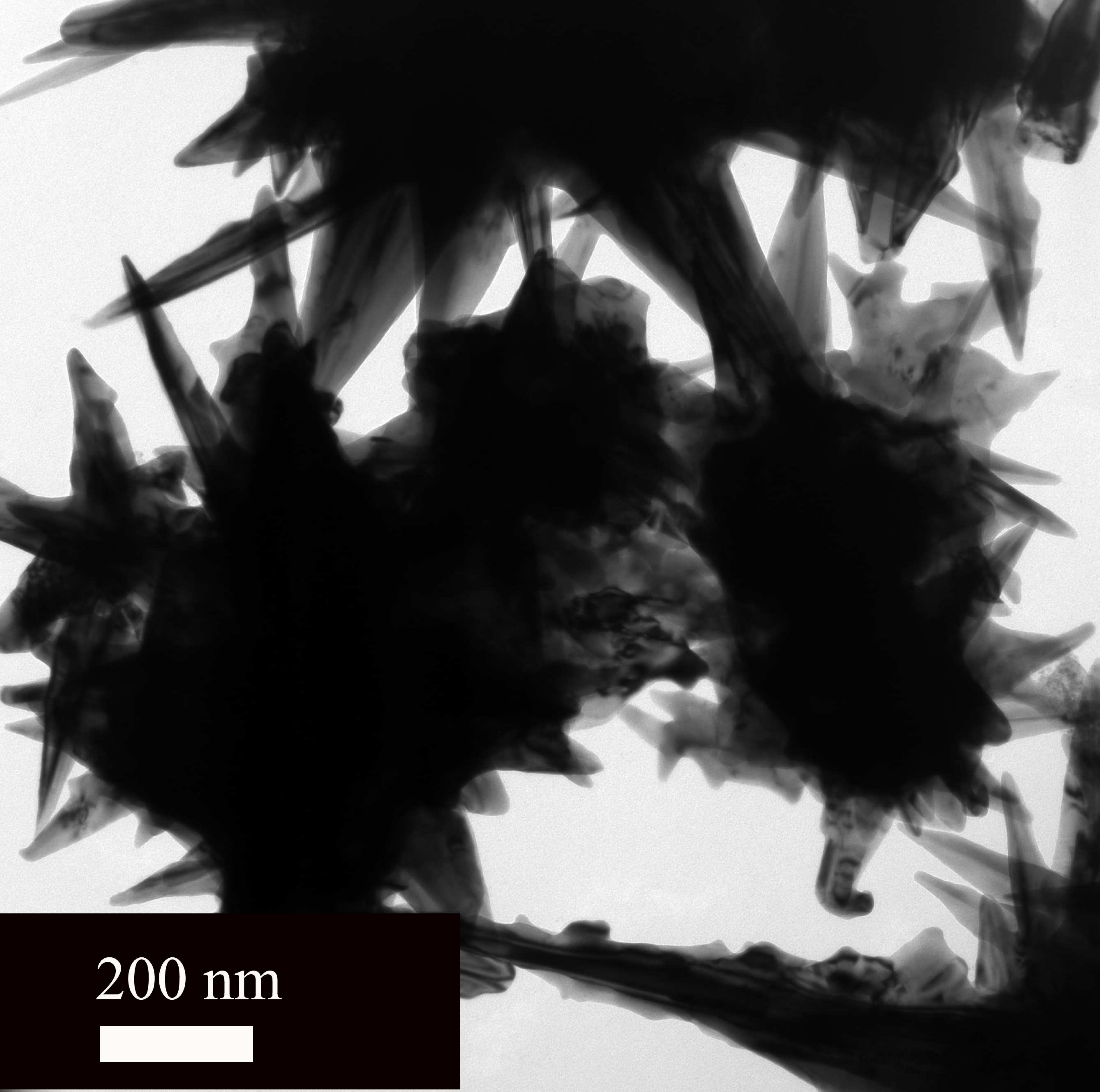}}%
\hspace{8pt}%
\subfigure[][]{%
\label{fig:2-b}%
\includegraphics[width=0.21\textwidth]{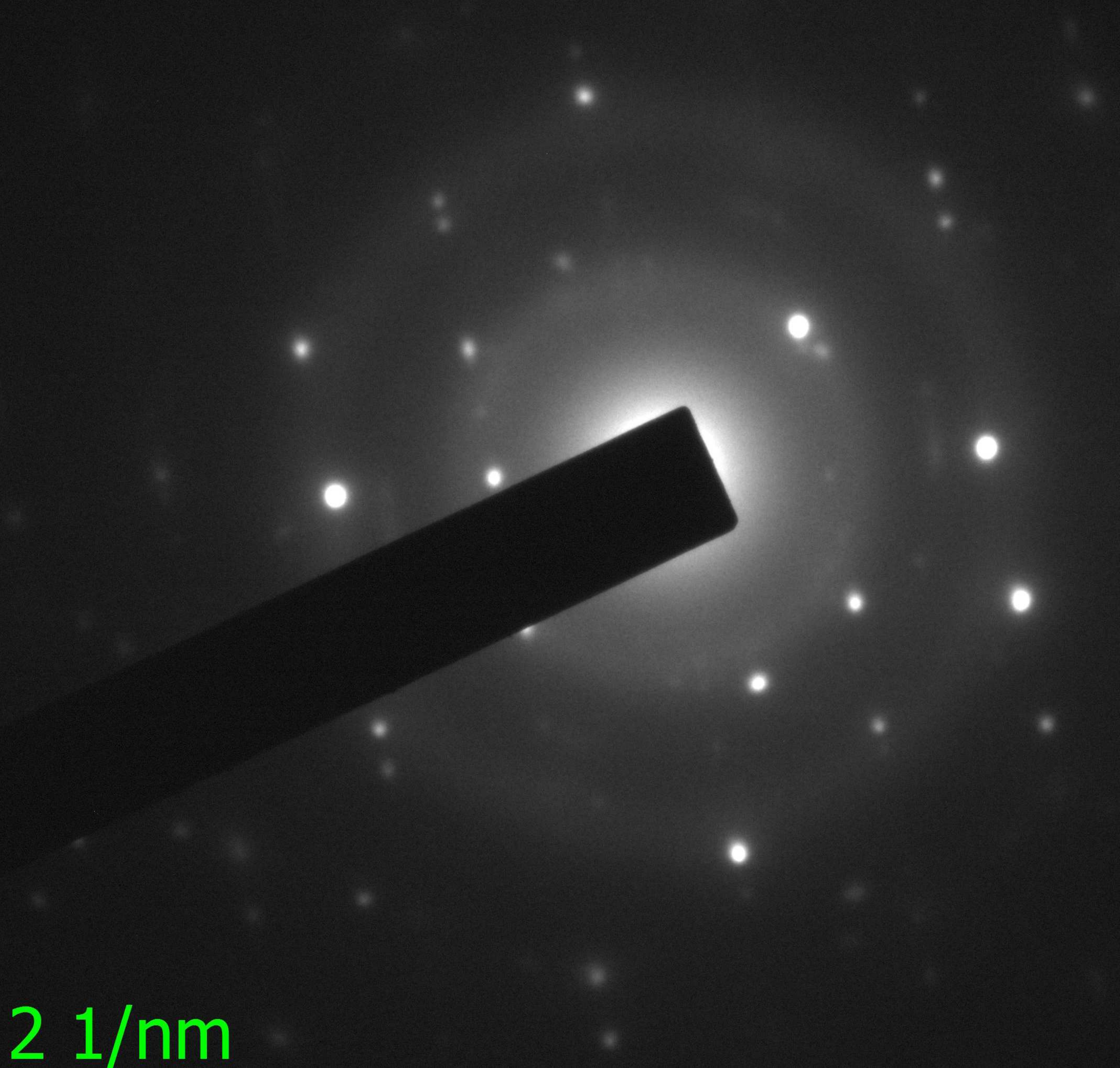}}%
\hspace{8pt}%
\subfigure[][]{%
\label{fig:2-c}%
\includegraphics[width=0.18\textwidth]{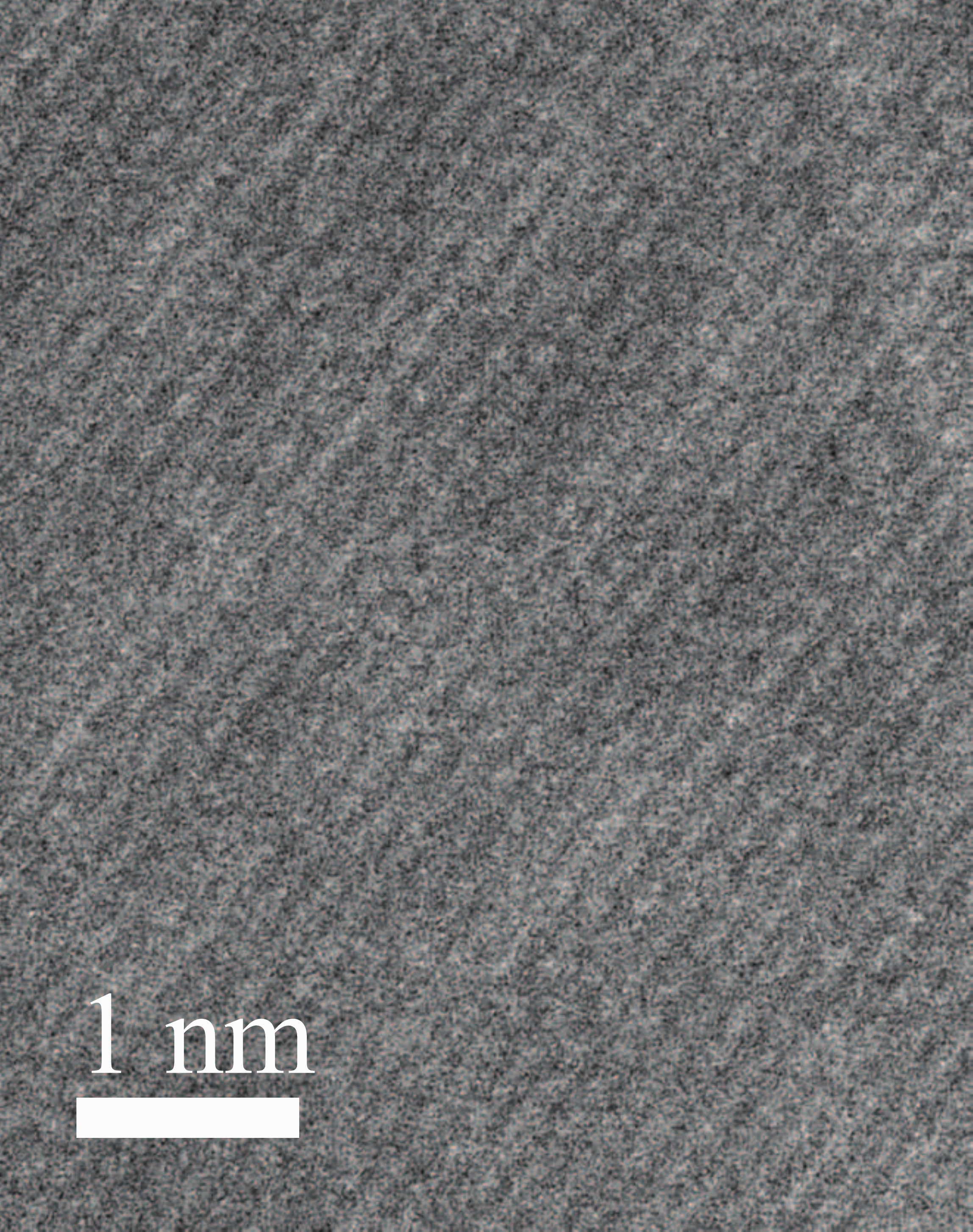}}%
\hspace{8pt}%
\subfigure[][]{%
\label{fig:2-d}%
\includegraphics[width=0.20\textwidth]{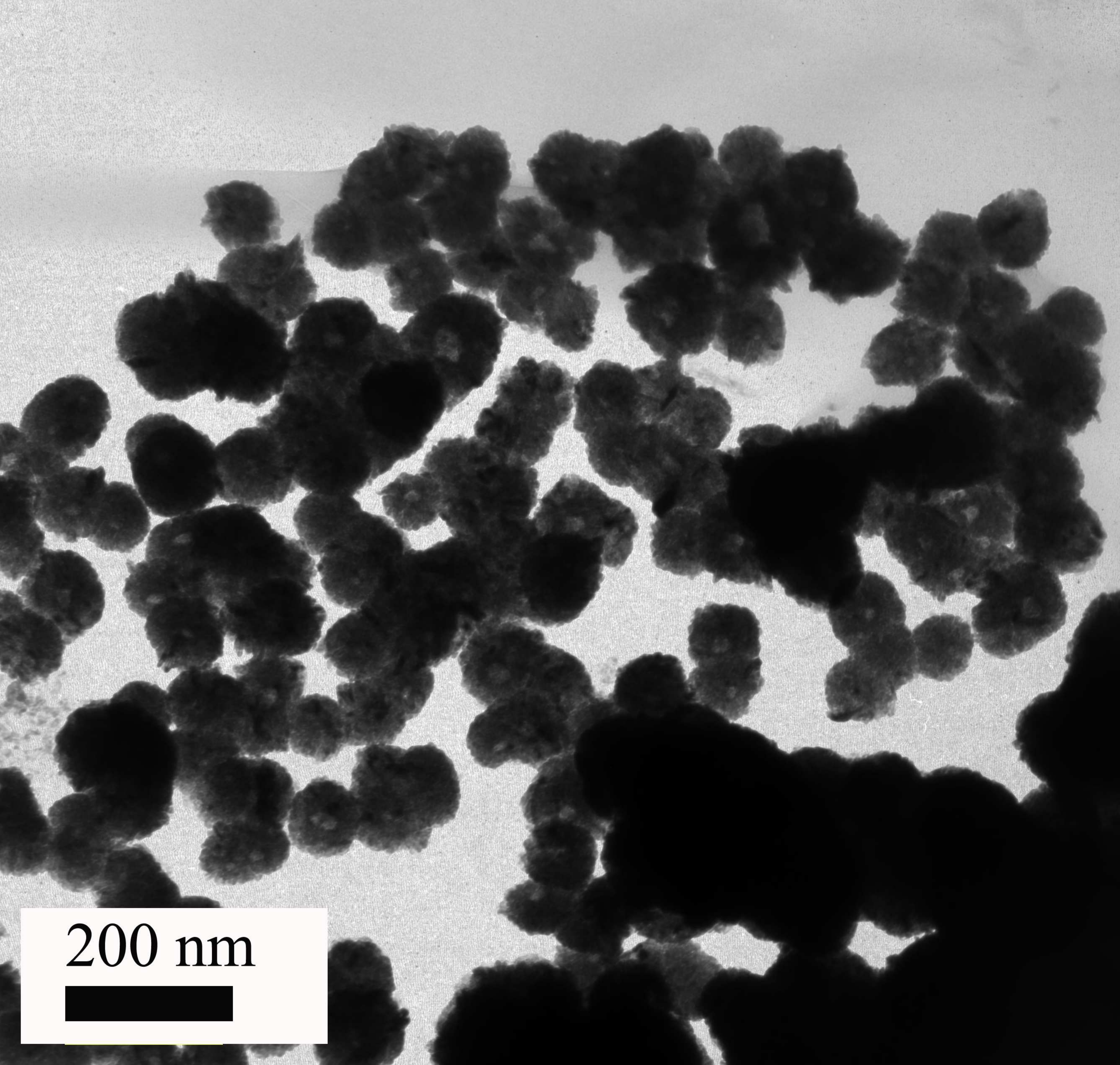}}%
\hspace{8pt}%
\subfigure[][]{%
\label{fig:2-e}%
\includegraphics[width=0.21\textwidth]{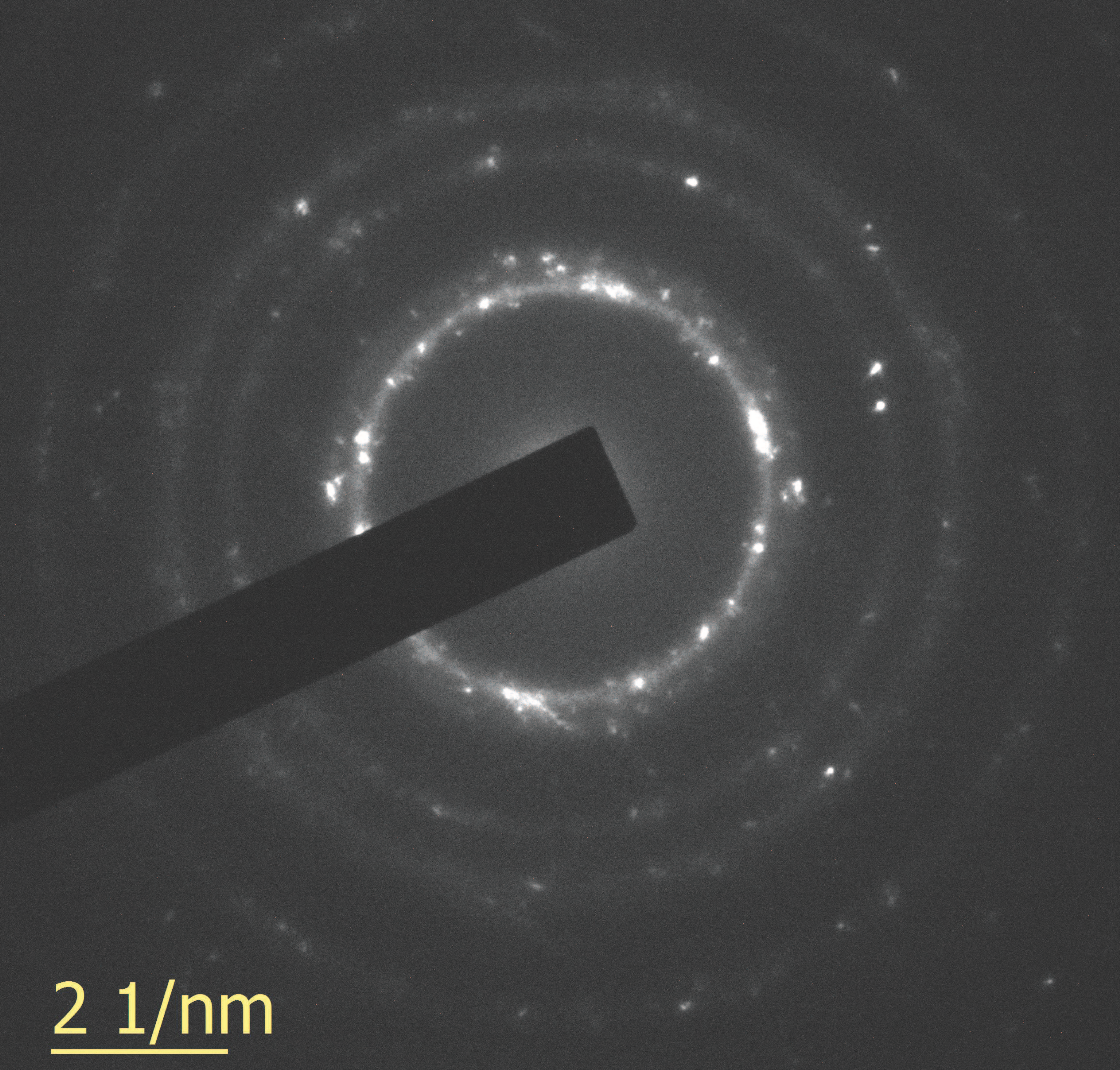}}%
\hspace{8pt}%
\subfigure[][]{%
\label{fig:2-f}%
\includegraphics[width=0.18\textwidth]{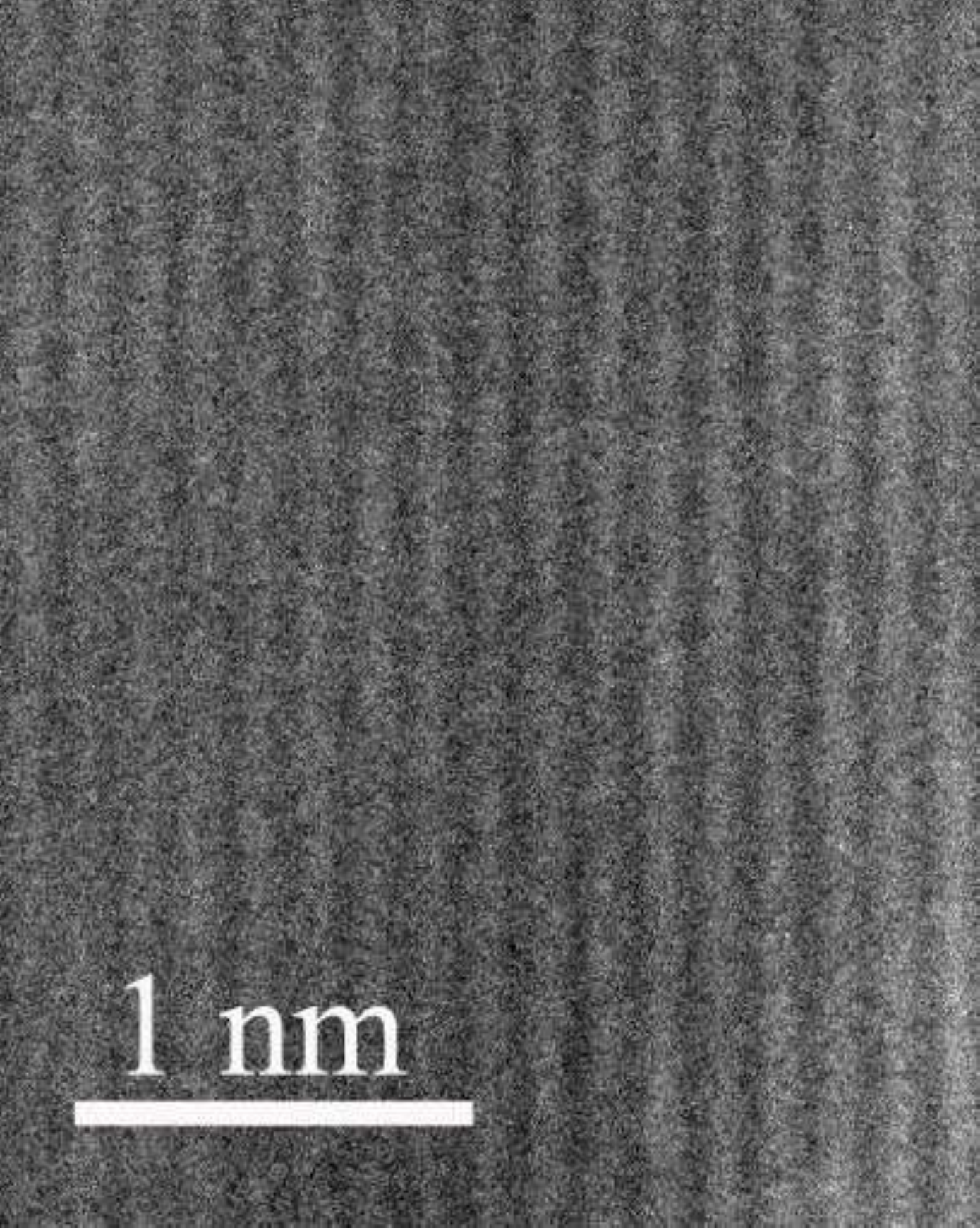}}%
\hspace{8pt}%
\subfigure[][]{%
\label{fig:2-g}%
\includegraphics[width=0.21\textwidth]{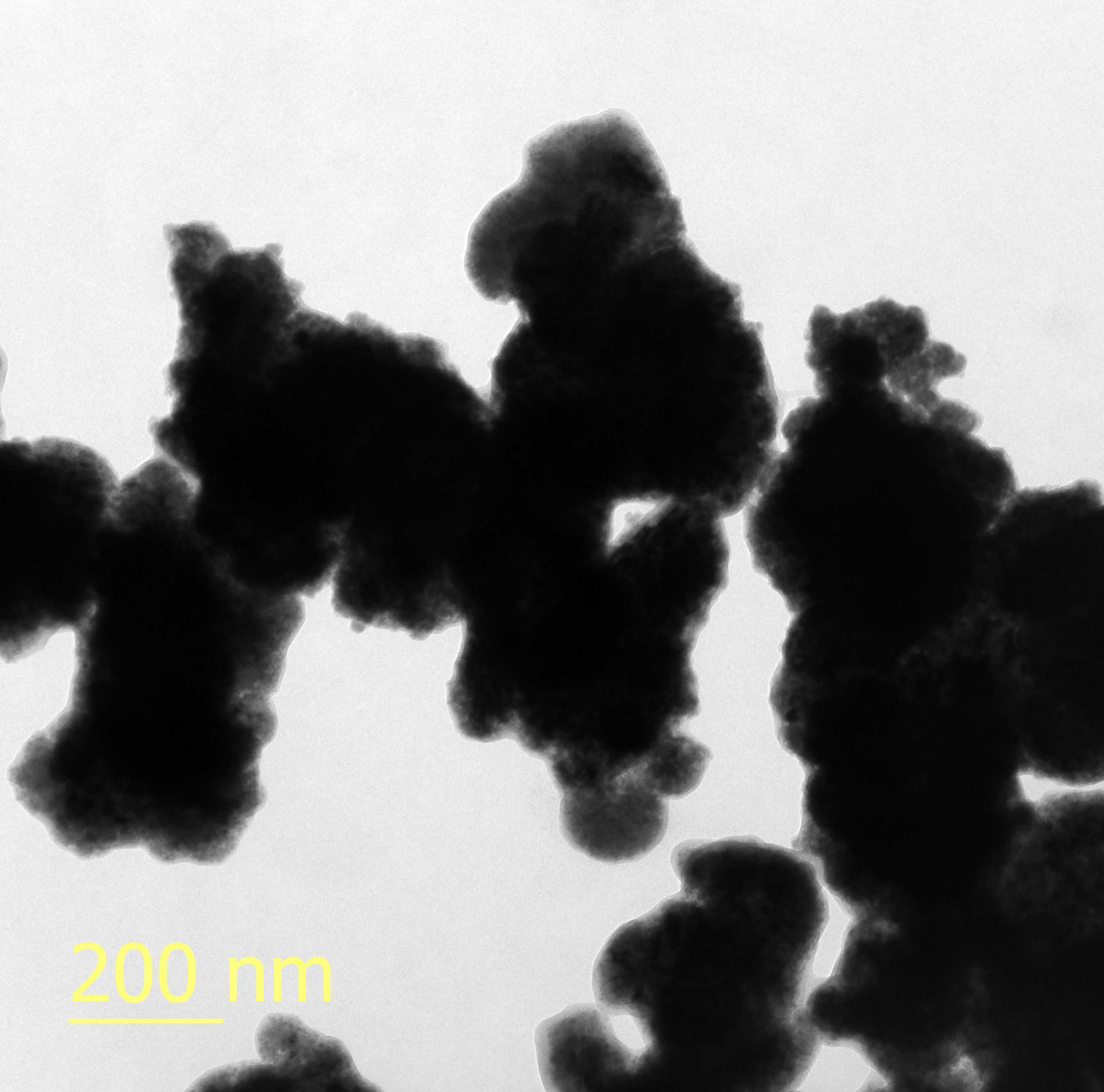}}%
\hspace{8pt}%
\subfigure[][]{%
\label{fig:2-h}%
\includegraphics[width=0.21\textwidth]{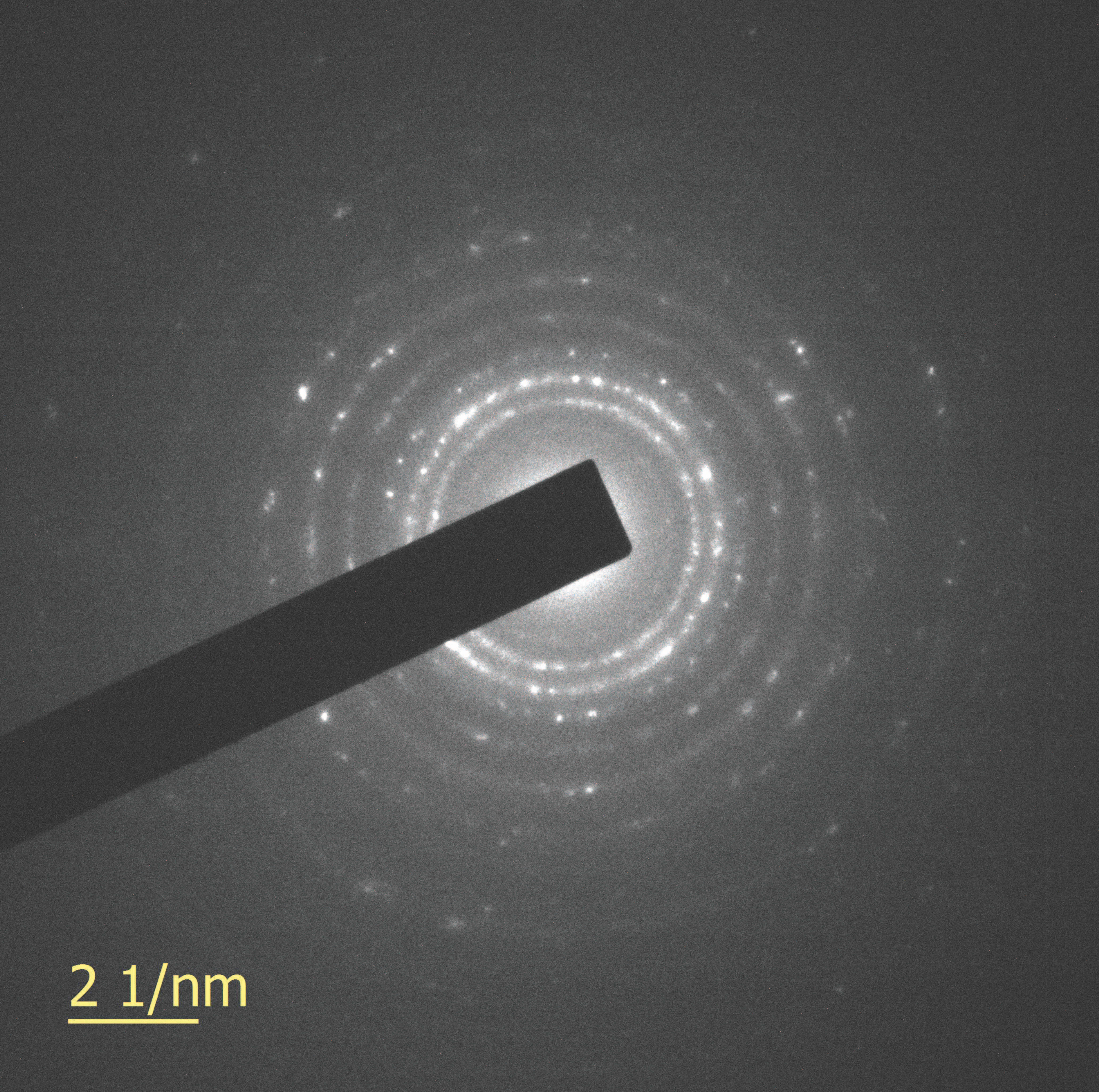}}%
\hspace{8pt}%
\subfigure[][]{%
\label{fig:2-i}%
\includegraphics[width=0.18\textwidth]{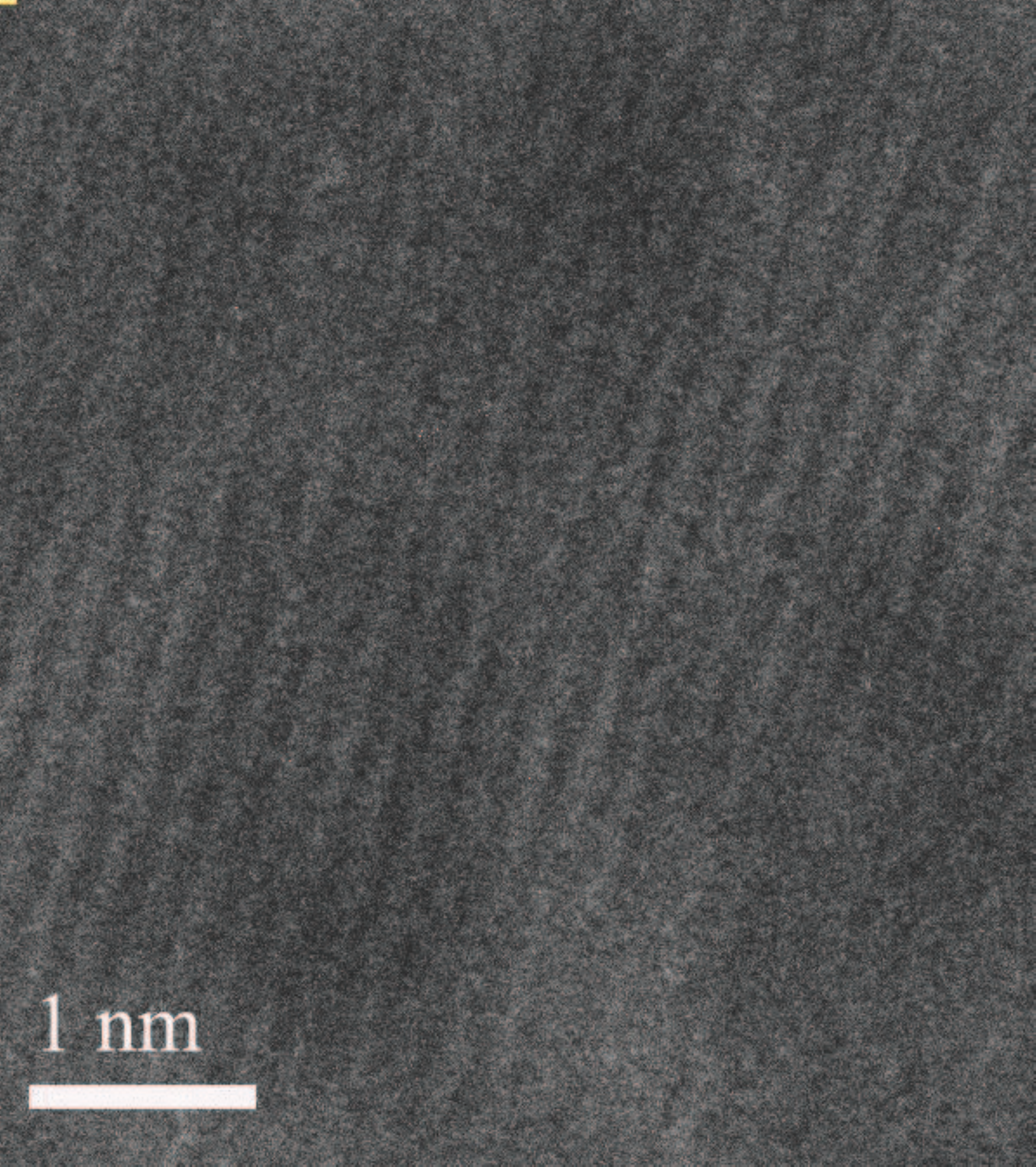}}%

\caption[]{HRTEM micrographs of Ni$_{1-x}$Rh$_x$ samples:

\subref{fig:2-a} the HRTEM image for $x$ = 0 of N6,

\subref{fig:2-b} the selected area diffraction pattern of $x$ = 0
of N6,

\subref{fig:2-c} a higher resolution image of $x$ = 0 of N6

showing the lattice planes,

\subref{fig:2-d} the HRTEM image for  $x$ = 0.27 of N6,

\subref{fig:2-e} the selected area diffraction pattern of $x$ =
0.27 of N6,

\subref{fig:2-f} a higher resolution image of $x$ = 0.27 of N6

showing the lattice planes,

\subref{fig:2-g} the HRTEM image for  $x$ = 0 of N25,

\subref{fig:2-h} the selected area diffraction pattern of $x$ = 0
of N25 and
\subref{fig:2-i} a higher resolution image of $x$ = 0
of N25 showing the lattice planes.}%
\label{fig:2}%
\end{figure}

Figure \ref{fig:1} shows the SEM images of pure Ni and
Ni$_{1-x}$Rh$_x$ alloy samples. In all the case we observe the
agglomeration of the nanoparticles which has taken place by
putting the drop on the aluminium sheet during the preparation of
nanoparticles for SEM. Except from Pure Ni  all other
Ni$_{1-x}$Rh$_x$ show formation of spherical nanoparticles with
30-40 nm. In case of pure Ni, for the two samples a flower type
pattern is observed. By a careful observation we found out that in
both the samples very small, spherical nanoparticles  of 30 -40 nm
agglomerated to form a flowery pattern. For more detailed
investigation the particle size as well as shape of the samples
were characterized using HRTEM. out of all the samples we have
selectively chosen three samples for HRTEM study. Fig. \ref
{fig:2} shows the HRTEM micrographs and the corresponding SAED
patterns of the Ni$_{1-x}$Rh$_x$ nanoparticles. From the HRTEM
images agglomeration of the nanopaticles is clearly observed.
Nanoalloys agglomerate in to bigger spherical bunch. This may be
due to the highly magnetic interaction between the
nanoparticles.\cite{ABC} The appearance of bright spots and
concentric rings in each SAED pattern show the solid evidence of
crystalline nature of the synthesized nanoalloys.

Fig. \ref{fig:2-c}, \ref{fig:2-f}, \ref{fig:2-i}  show the lattice
fringes of Pure Ni that means x = 0 of N 6, N 25 and
Ni$_{0.73}$Rh$_{0.27}$ of N6 nanoparticles respectively. All the
three samples exhibit straight edges and clear lattice fringes.
The average inner plane distances (d-spacing) for adjacent fringes
for pure x = 0 of  N 6 and N 25 nanoparticle were found to be
1.74$A^0$ and 2.1 $A^0$. For former average inner plane distances
(d-spacing) for adjacent fringes is significantly close to  fcc Ni
(200) and for latter, it is close fcc Ni (111) calculated from the
XRD pattern of both the samples by PCW. In case of
Ni$_{0.73}$Rh$_{0.27}$, the average inner plane distances
(d-spacing) for adjacent fringes was measured to be 2.2 $A^0$,
which coincides with the lattice spacing d  of the (111) planes of
Ni$_{0.73}$Rh$_{0.27}$ determined from the XRD by PCW. From this
it is  confirmed  that the synthesized nanoparticles are
completely in alloy form.

\subsection{XRD}

\begin{figure}%
\centering
\subfigure[][]{%
\label{fig:3-a}%
\includegraphics[width=0.45 \textwidth]{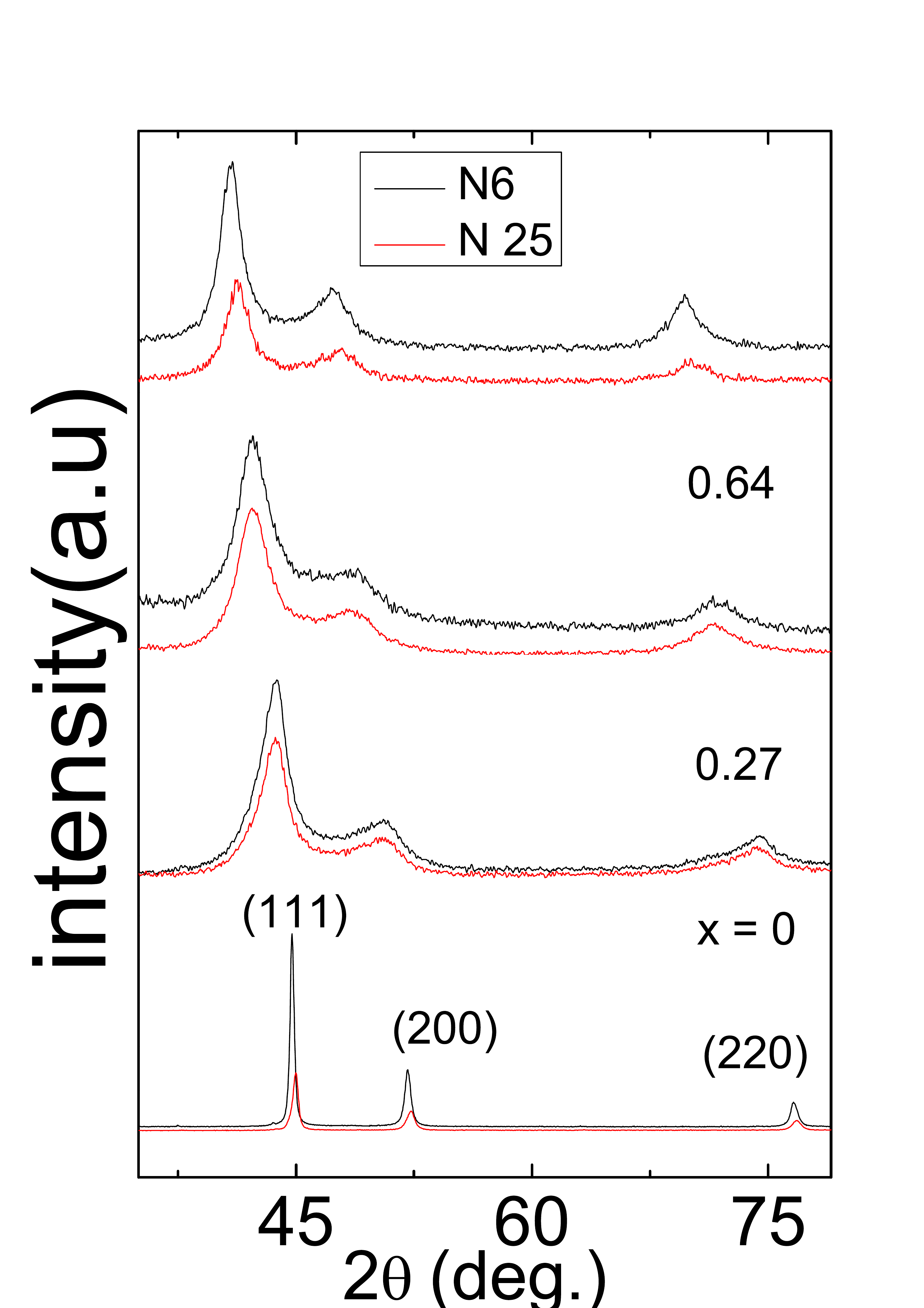}}%
\hspace{8pt}%
\subfigure[][]{%
\label{fig:3-b}%
\includegraphics[width=0.45\textwidth]{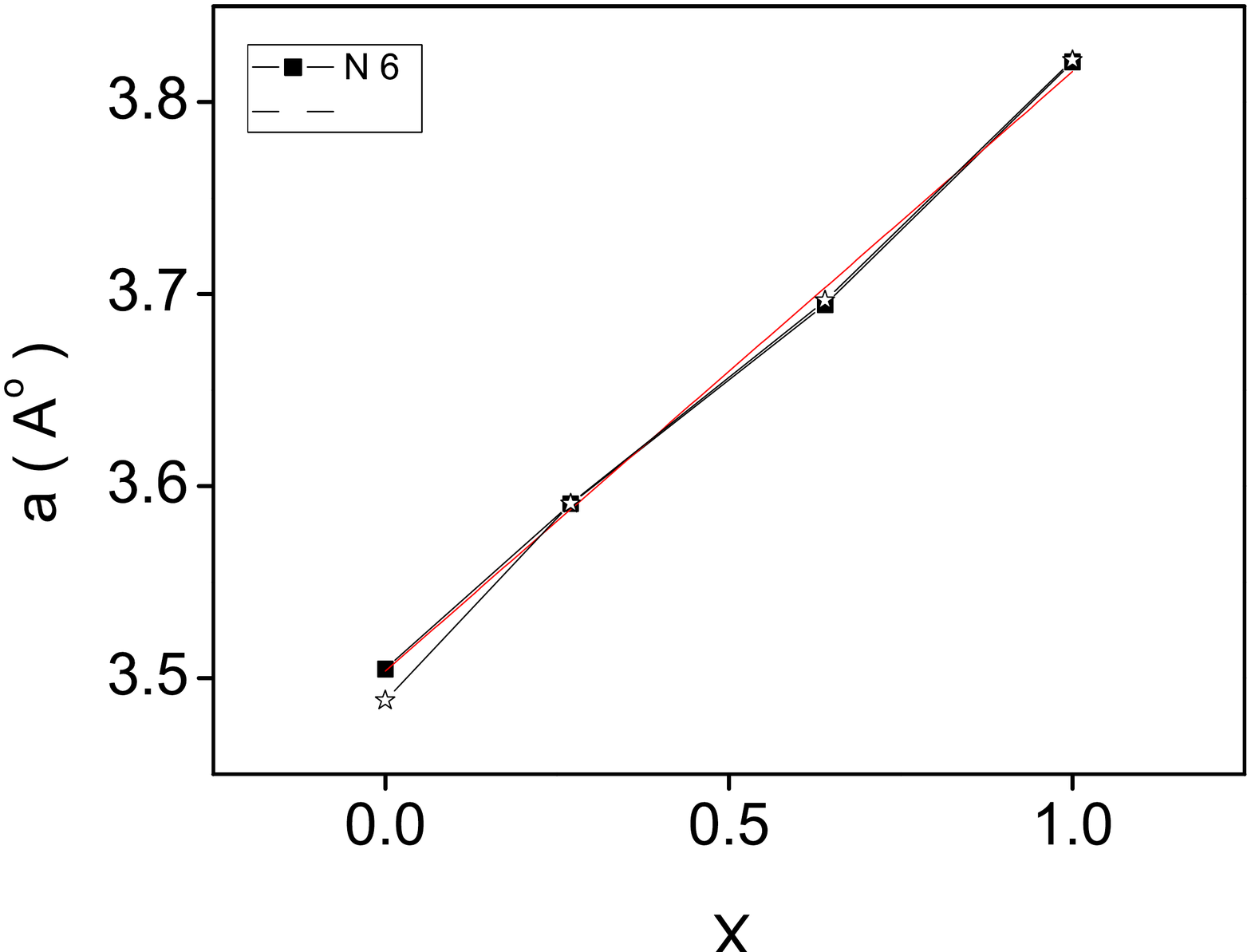}}%
\hspace{8pt}%

\caption[]{(a) XRD patterns of Ni$_{1-x}$Rh$_x$ samples. (b) Variation of lattice constant $a$ with $x$.}
\label{fig:3}%
\end{figure}
Figure\ref{fig:3-a} shows typical XRD spectra of all the
synthesized pure Ni and  Ni$_{1-x}$Rh$_x$ alloy nanoparticles. For
the comparison the XRD pattern of pure Rh nanoparticles prepared
in the same manner are also included in the figure. In the XRD
spectrum of pure Ni and pure Rh three sharp strong reflection
peaks at $2\Theta$ values of $44.78^o$, $52.14^o$, and $76.62^o$,
and $41.05^o$, $47.524^o$ and $69.688^o$ appears respectively.
According to JCPDS-ICDD powder diffraction database these peaks
are correspond to the (111), (200), and (220) planes of the fcc
crystallographic structure of Ni and Rh respectively.However the
X-ray diffraction peaks of the Ni$_{1-x}$Rh$_x$ nanoalloys matches
the (111), (200), and (220) characteristics peaks  of a Rh fcc
structure but slightly shifted to higher $2\theta$ values. There
were also no observable lines in the XRD spectra corresponding to
those of pure Rh or Pure Ni. If the homogeneous solid solution of
Rh-Ni was not formed, then the peaks of pure Rh or Pure Ni
 would have observed in the Spectra. The absence of the
Pure Ni and Pure Rh peaks in the XRD patterns of all the alloy
nanoparticles suggests a complete alloying of Ni and Rh for all x
values under study. The shift in  $2\theta$ in curve a corresponds
to a increase in the lattice constant due to the incorporation of
Rh atoms in the Ni. Further, Fig. \ref{fig:3-b} shows the
variation of lattice constants determined from all XRD patterns
with x. The lattice parameter increases with increase of the Rh
concentration is same as in case of bulk Ni$_{1-x}$Rh$_x$ alloy.
\cite{Nash} From the fig. it is confirmed that nanoalloys of this
also obey Vegard's law.

 \subsection{XPS}

\begin{figure}%
\centering
\subfigure[][]{%
\label{fig:4-a}%
\includegraphics[width=0.45\textwidth]{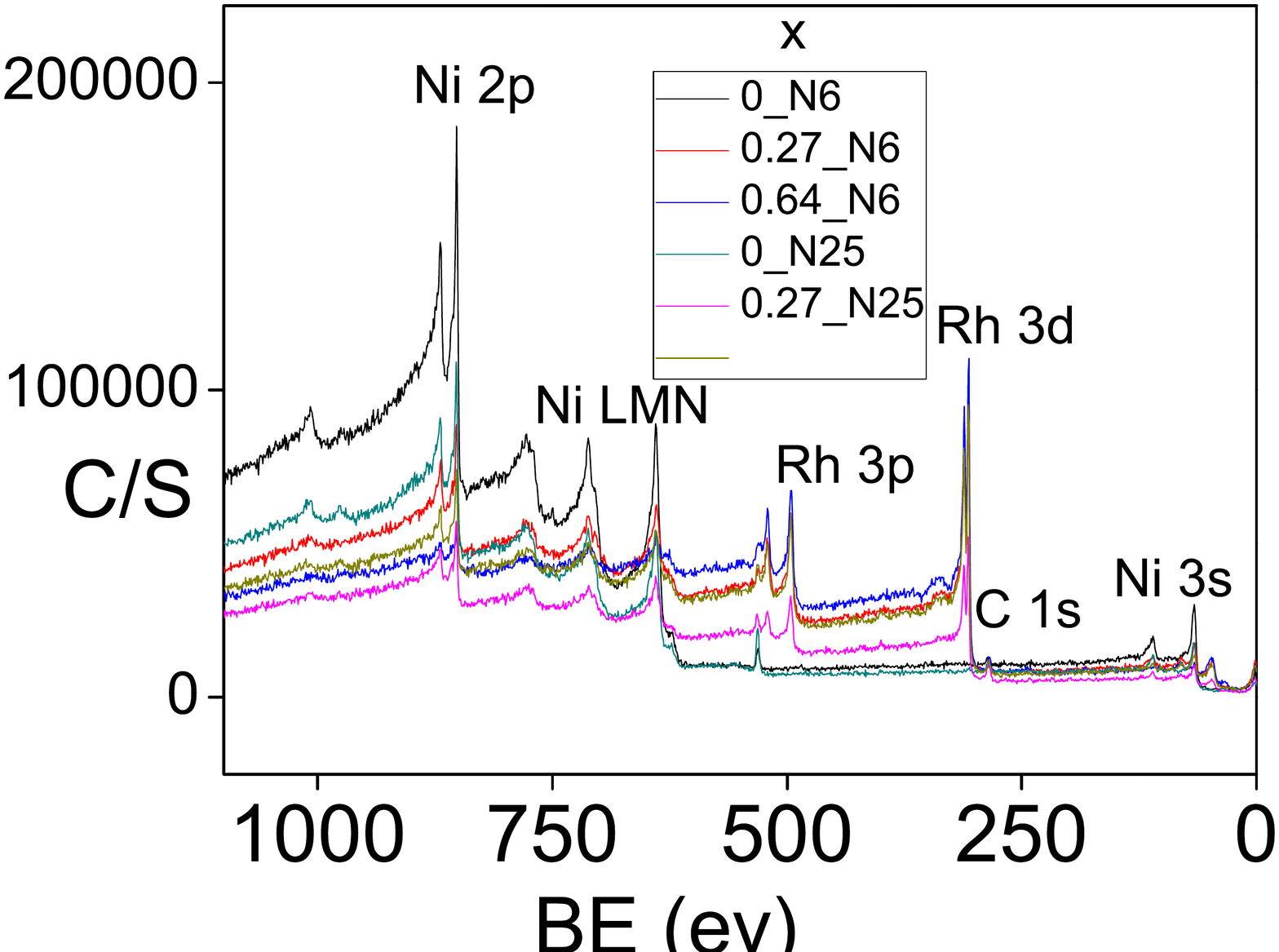}}%
\hspace{8pt}%
\subfigure[][]{%
\label{fig:4-b}%
\includegraphics[width=0.35\textwidth]{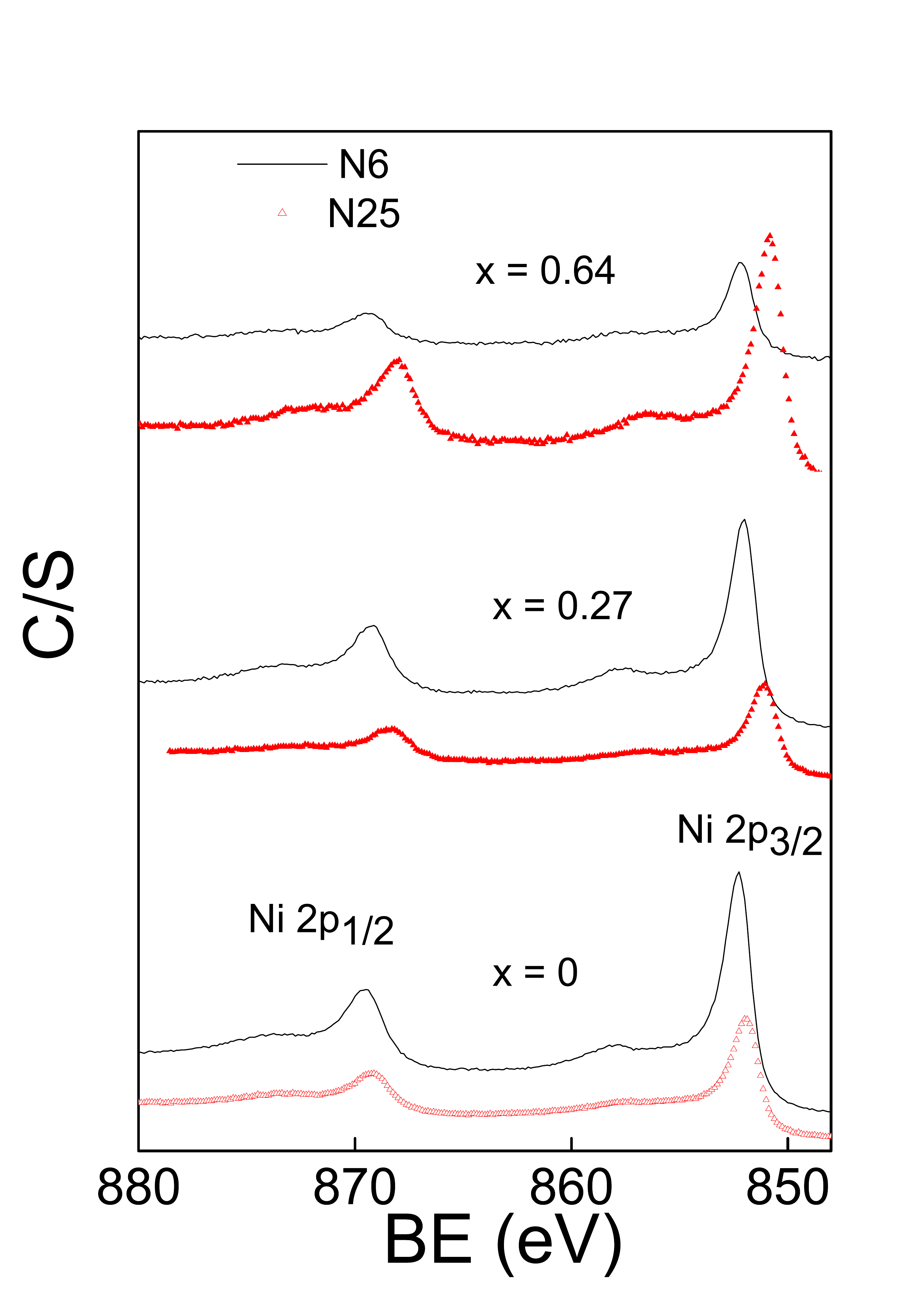}}%
\hspace{8pt}%
\subfigure[][]{%
\label{fig:4-c}%
\includegraphics[width=0.35\textwidth]{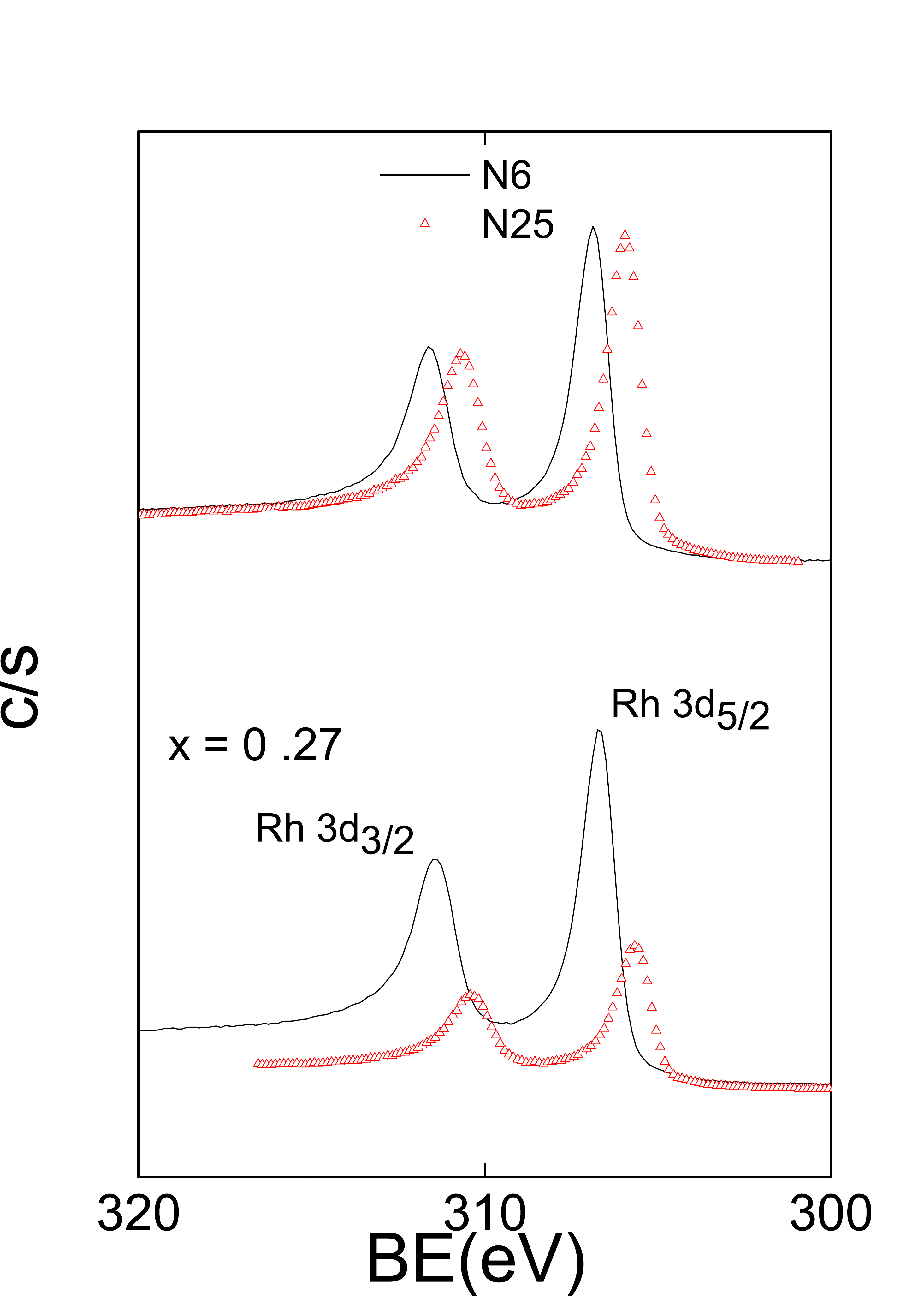}}%
\hspace{8pt}%

\caption[]{(a) Survey XPS spectra of Ni$_{1-x}$Rh$_x$ samples. (b) High-resolution XPS spectra in the Ni 2p region. (c) High-resolution XPS spectra in the Rh 3d region.}
\label{fig:4}%
\end{figure}

 The metallic Ni$_{1-x}$Rh$_x$ alloy nanoparticle  were also studied by X-ray
photoelectron spectroscopy (XPS). The survey scans of XPS spectra
for Ni$_{1-x}$Rh$_x$ alloy displayed in figure\ref{fig:4-a} show
pronounced peaks Rh and Ni. The c 1s peak which appears at 284.5
ev and O 1s at 531.0 ev  are absent for N6 samples. However for N
25 samples the c 1s and O 1s are present. The intensity and height
of the peaks are negligible which appears due to the unavoidable
hydrocarbon and oxygen in the atmosphere. In order to find the
alloying and the different oxidation states of Rh and Ni, the
nanoalloys were examined by high-resolution XPS in narrow range of
Rh 3d peaks and Ni 2p peaks  regions in more detail. Figure
\ref{fig:4-b} and figure \ref{fig:4-c}shows the high-resolution
XPS spectra for all samples in the Ni 2p region and  Rh 3d region
respectably. In case of N 25 samples one would expect the presence
of oxygen peak may contribute for the formation of oxides. One
would expect a peak between at 853.7,  855.6 eV  and  308.5 eV  if
the sample contains any  NiO or $ Ni(OH)_2$ or $ {Rh}_2O_3$ This
gives a further confirmation of synthesized samples are oxide
free. It is to be noted that in case of bulk metallic Ni, the Ni
$2p_{3/2}$ appears at 852.7 eV, we observe a binding energy shift
of about +0.36 eV and +0.2 eV in pure Ni  N6 and N 25 respectively
with respect to bulk however such shifts are expected in
nanoparticles. \cite{Liu2009}. The Ni 2p peaks of the alloys
shifts towards lower binding energy with the increase of Rh
concentration. In case Rh 3d peaks, the peaks are shifted towards
higher binding energy with the increase of Rh concentration. This
is because between Ni and Rh, Rh is the more electronegative so
upon alloying with Ni, a charge transfer from the Ni site to the
Rh site is expected. So the Binding energy of Ni is shifted toward
lower side and Rh towards the higher sides.

 \subsection{Magnetization}

\begin{figure}%
\centering
\subfigure[][]{%
\label{fig:5-a}%
\includegraphics[width=0.45\textwidth]{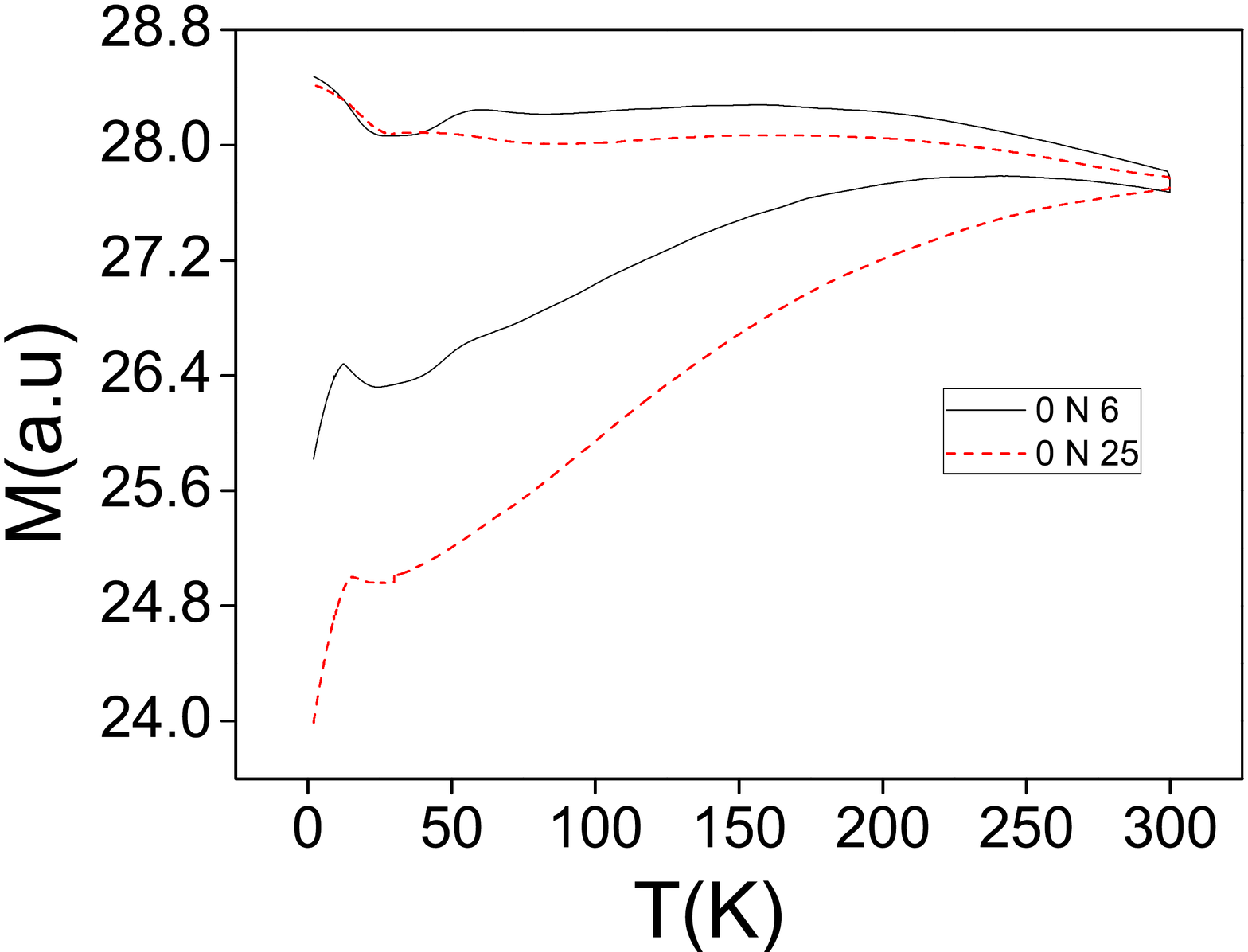}}%
\hspace{8pt}%
\subfigure[][]{%
\label{fig:5-b}%
\includegraphics[width=0.45\textwidth]{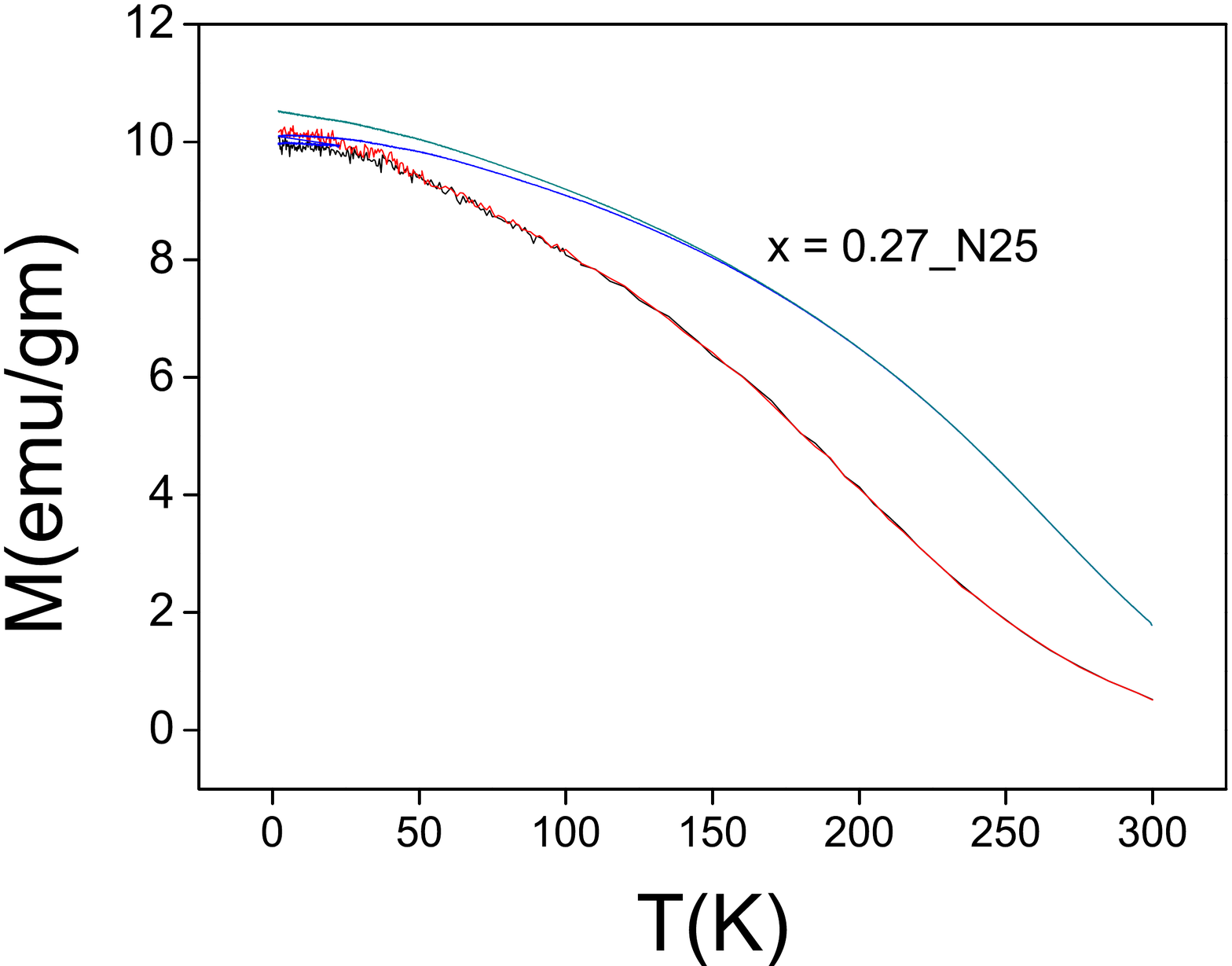}}%
\hspace{8pt}%
\subfigure[][]{%
\label{fig:5-c}%
\includegraphics[width=0.45\textwidth]{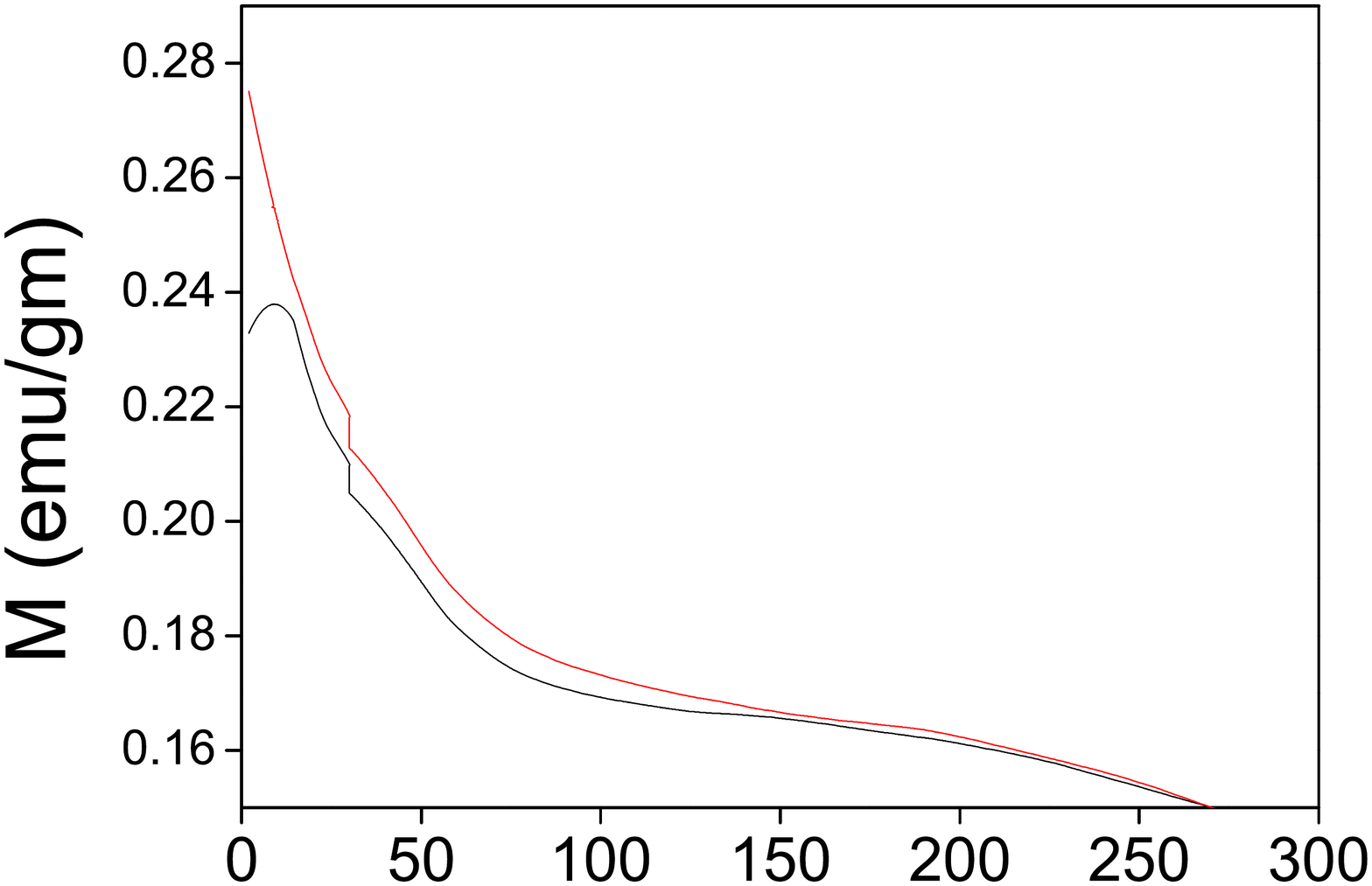}}%
\hspace{8pt}%
\subfigure[][]{%
\label{fig:5-d}%
\includegraphics[width=0.45\textwidth]{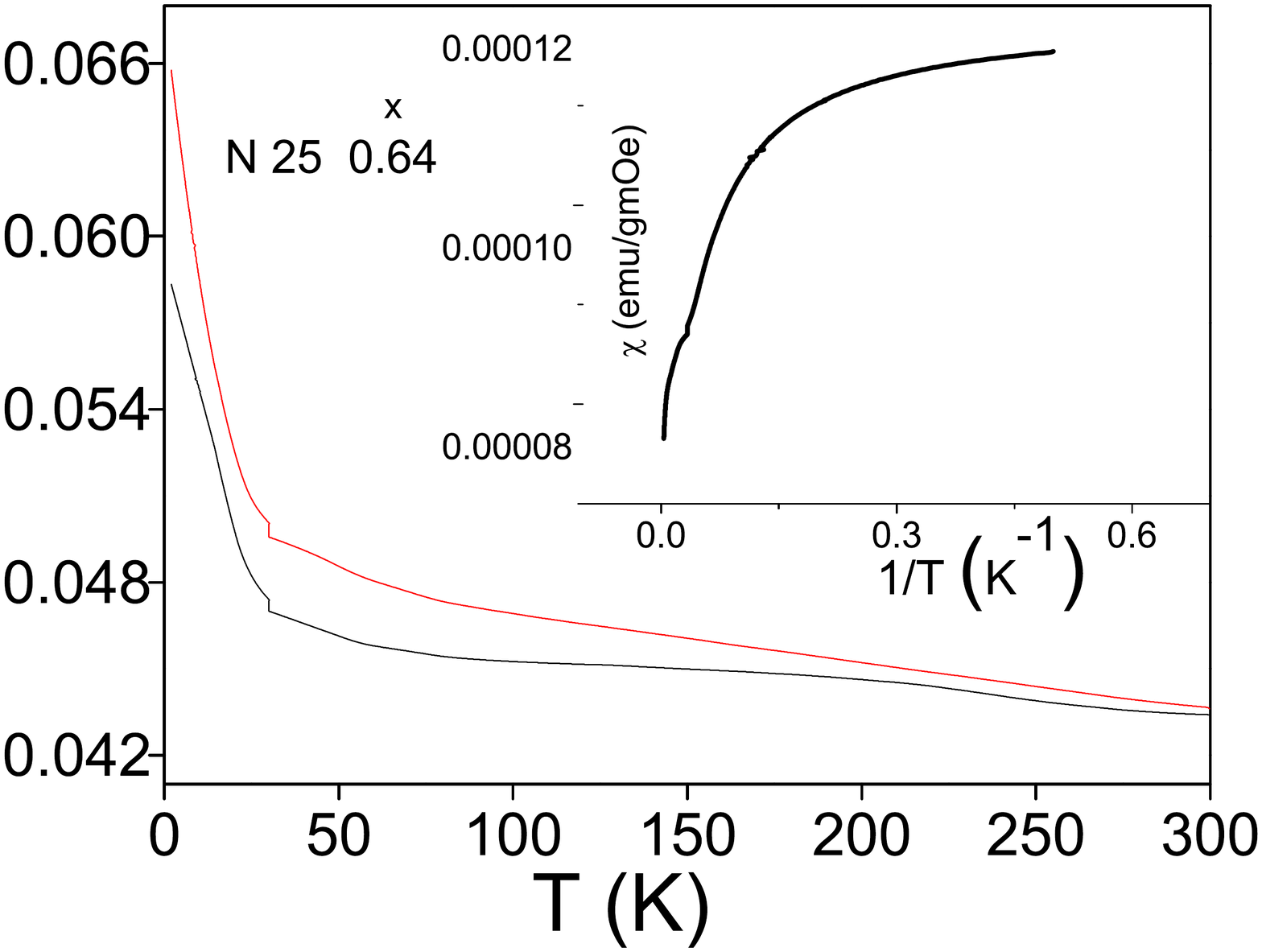}}%
\hspace{8pt}%

\caption[]{FC and ZFC magnetization versus temperature (M -T)
plots of Ni$_{1-x}$Rh$_x$ samples at 500 Oe field:

\subref{fig:5-a} M-T plot for  $x$ = 0of N6 and N 25,

\subref{fig:5-b} M-T plot for  $x$ = 0.27 of N6 and N 25,

\subref{fig:5-c} M-T plot for  $x$ = 0.64 of N6,

\subref{fig:5-d} M-T plot for  $x$ = 0.64 of N25,and the inset
show the variation of $\chi $ vs. $1/T$.}%
\label{fig:5}%
\end{figure}

\begin{figure}%
\centering
\subfigure[][]{%
\label{fig:6-a}%
\includegraphics[width=0.45\textwidth]{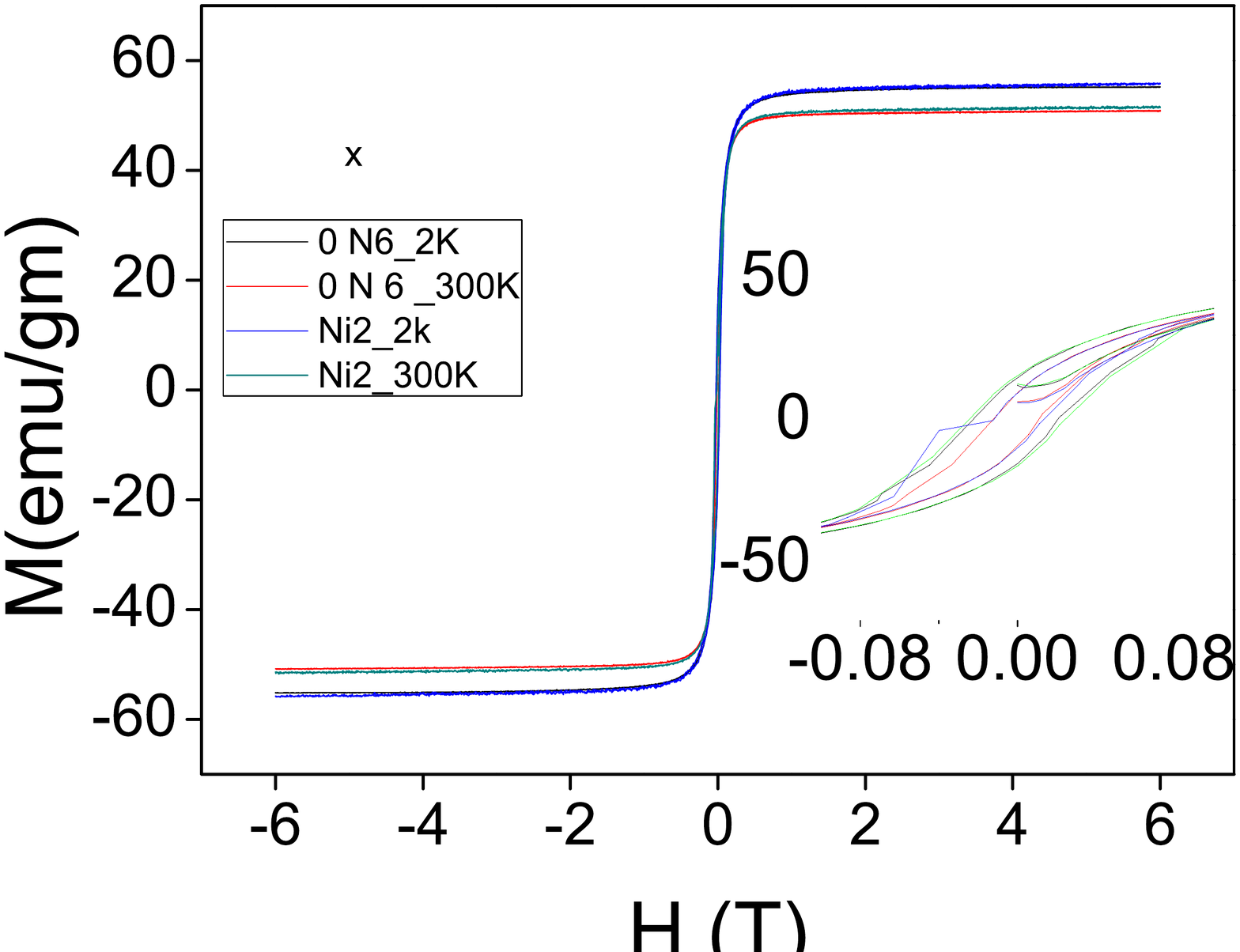}}%
\hspace{8pt}%
\subfigure[][]{%
\label{fig:6-b}%
\includegraphics[width=0.45\textwidth]{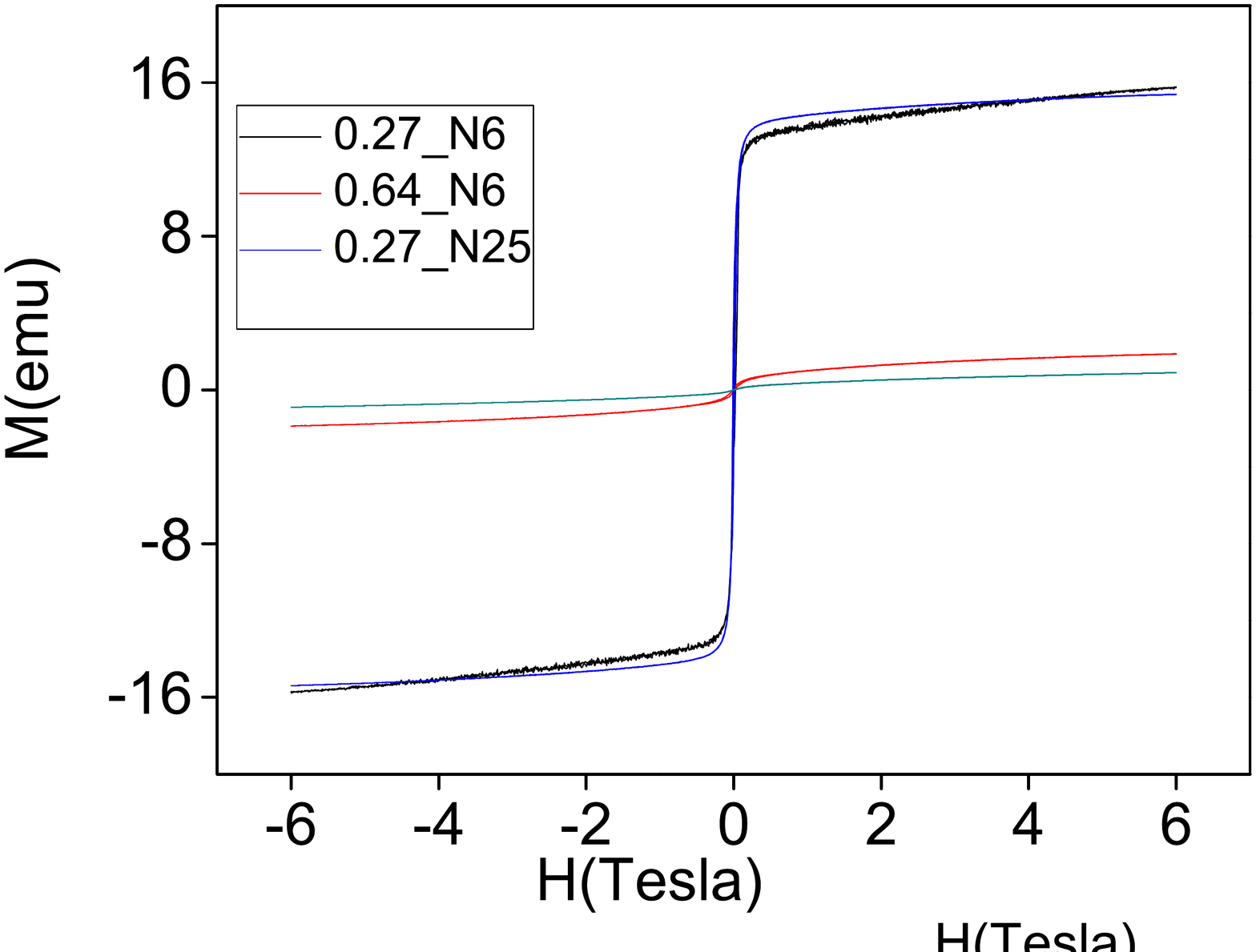}}%
\hspace{8pt}%
\subfigure[][]{%
\label{fig:6-c}%
\includegraphics[width=0.45\textwidth]{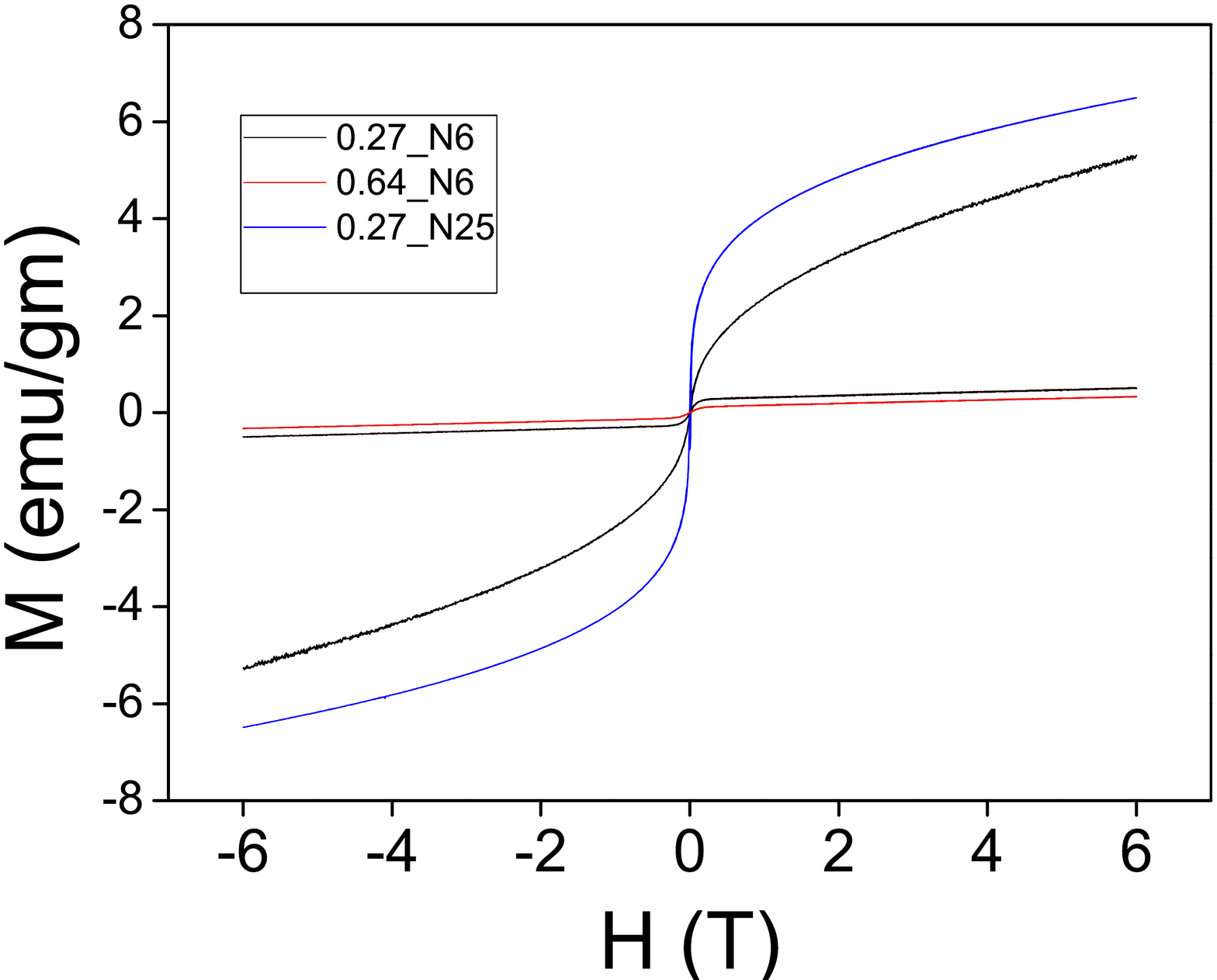}}%
\hspace{8pt}%

\caption[]{ M - H plots of Ni$_{1-x}$Rh$_x$ samples:

\subref{fig:6-a} M-H plot for  $x$ = 0 of N6 and N 25 at 2 k and
300 K, \subref{fig:6-b} M-H plot for  $x$ = 0.27 and $x$ = 0.64
of N6 and N 25 at 2 K, \subref{fig:6-c} M-H plot for  $x$ = 0.27
and $x$ = 0.64   of N6
and N 25 at 300 K.}%
\label{fig:6}%
\end{figure}

Figure \ref{fig:5} represents the variation of magnetization with
temperature, measured under an applied field of 500 Oe in the
standard zero-field-cooled (ZFC) and field-cooled (FC) protocols.
We will discuss the ZFC -FC magnetization study of each
synthesized concentration separately one by one.

For Pure Ni nanpparticle the ZFC and FC curves show some
interesting features worth mentioning:  The ZFC and FC curves show
some interesting features worth mentioning:(1)The irreversibility
in the ZFC and FC curves present through the whole measuring
temperature range in both N6 and N 25 samples. (2) Taking in the
consideration of the ZFC curve in both the samples show  quiet
similar type of behavior. We observe two peaks one sharp and
strong peak in low temperature region and another broad peak in
the higher temperature region. The only difference in the samples
is the lower temperature peak, which is very intense in sample N6
in comparison to sample N25. In sample N6 going down from 300 K in
the temperature axis the magnetization first increases followed by
a broad peak  at around 241 K  then decreases up to 22 K and then
increases showing a sharp peak at 12 K and finally decreases down
to low temperature. In case of sample N25 similar behavior is
observed however the high temperature peak is more broad than N 6
sample and the low temperature peak appears at 14 K.
 Here we try to explain the above features one by one.
  The peak in ZFC  and irreversibility between  ZFC and Fc curve are the characteristic
features of SPM nanoparticles. \cite{Bedanta 2009} In order to
understand the existence of two peaks in the ZFC curve in both the
samples, we propose that the sample is consisting of two types of
particles the small particles present at the surface and bigger
larger particles inside. The big ones have SPM behavior and show a
blocking temperature $T_B$  in the higher temperature side and the
smaller particle have a blocking temperature at very low $T_S$ $<$
15 K. \cite{chen 2005} The broadening in the peak is due the
agglomerations. In sample N 25 the size of particles is more than
sample N 6 as confirmed from SEM images, due to why the $T_B$ is
more broaden in former one. we can also say that the size
distribution of the bigger SPM nanoparticles are large in
comparison to the small surface particles. Now we will discuss
features in the FC curve in the two samples. In Sample N6 by
magnetization value increases with the decrease of temperature
from the 300 K up to 60 K  then increases followed by a maxima in
the FC at 57 K. Below 57 K the magnetization decreases showing a
dip  or minima at 32 K and by further decrease of temperature
Magnetization increases without saturating. The monotonic increase
of magnetization in FC below $T_B$ with the temperature is the one
of the key feature of SPM nanoparticles. However in sample N6 the
increase of Magnetization with T up to 60 k show little bit flat
this is due to  inter particle inter action present in the between
SPM nanoparticles. Due to the interaction between the SPM
nanoparticles the nanopaticles freezes at low temperatures which
is appeared as a hump in FC curve at 57 K corresponding to the
glass temperature of the system below which the system shows a
glassy behavior. \cite{Suzuki 2009} However minimum in
Magnetization upon cooling is only observed in  super spin glass
system (SSG). \cite{Bedanta 2009, Sasaki 2005, Petracic 2006} Well
established evidences of presence of  a superspin glass (SSG)
below a well-defined glass temperature, $T_g$  in various systems
has been studied.\cite {Djurberg, Jonsson,Petracic} The decrease
in the magnetization below 57 K is due to the collective frizzing
of the spins. \cite{Suzuki 2009, Khurshid 2015} Sample N 25 show
similar type of behavior only the hump which is referred as the
spin glass temperature and the minima in the FC upon cooling which
is characteristics feature of SSG system is less intense. From
this argument we can state that in both the samples, the SSG
behavior is observed. But now the question arises what is the
cause of SSG type of behavior in our system. The glassy behavior
in the nano systems are due to the inter particle interaction,
anisotropy and the surface effect. SSG system is nothing but
analogous to spin glass state to the bulk. The only thing SSG is
the collective freezing of interacting spins in the nanosystem. In
our case the SSG phase arises due to the high inter particle
interaction between the SPM nanoparticles also observed in SEM and
HRTEM images and the interaction between the small surface
particles and larger SPM particles which cause a collective
frizzing of spin at low temperature. In Ni$_{0.73}$Rh$_0.27$ alloy
of N6 and N25  FC and ZFC  show similar behavior with Temperature.
However in former ZFC and FC are identical and coincides but in
latter the ZFC and FC bifurcates from 200 K. The variation of
Magnetization in both the samples show a typical ferromagnetic
type behavior with  a high transition temperature$T_C$. In
Ni$_{0.36}$Rh$_0.64$ of N6 FC and the ZFC curves stat splitting
from room temperature and the ZFC exhibiting a maximum near 9.1  K
which is the so called blocking temperature($T_B$) of the single
domain particles characterized by superparamegnetism region above
it and a blocked ferro region below it. The ZFC and FC curve does
not show any overlapping up to 300 K which means the largest
particle blocked at 300 K, the highest temperature limit of our
 study. Below $T_B$ Magnetization value for the FC increases monotonically
 with the decrease of temperature which is the significant
 property of  non interacting, single domain particles. However for N25 (x = 0.64), a strong increase of magnetization with the decrease
 of temperature  is observed which indicates that it becomes more
 paramagnetic than N6 sample. Furthermore, the M-T curve shows a irreversible ZFC/FC cycle, with the sharp upward curve also indicative of
 paramagnetism. However the deviation from the straight line behavior of  susceptibility with inverse temperature shown in figure
  inset of \ref{fig:5-d} confirms that the sample is not purely paramagnetic. A ferromagnetic interaction is present along with the
 paramagnetism. By a careful observation, we found there is a shoulder like feature appeared at 30 K in both ZFC and FC for both N6 and N 25
 samples. We named the temperature as $T_g$, the spin glass temperature, which is more pronounced in N25 than the other
 one. We will try to understand the magnetism of the origin of spin
 glass in the alloys. Normally the spin glass is observed in a system due to random magnetic interactions and frustration of
 spin. The magnetism in case of Ni$_{1-x}$Rh$_x$ alloy is due to interacting
 localized cluster moments similar to that of their bulk. When we doped Ni with Rh, the Rh took the random
 lattice position so the  exchange interaction between the Ni atoms breaks down but some places have Ni clusters, giving rise to the ferromagnetism.
 By further increasing the Rh concentration the no of Ni atom nearest neighbor to a given Ni atom decreases which given rise to the decrease
 in moment. Like the bulk here also we observe spin glass phase
 below the ferromagnetic phase. For bulk, the alloy show a short range interaction due to which Ni prefers being surrounded by Rh atoms. The spin glass phase
 in bulk arising by controlling the short range interaction by making the system random by some cold work.\cite{Carnegie93} However  in case nano,  the systems are a complete homogeneous alloy of Rh and Ni without any short range interaction. The origin of spin glass phase could be
due to the same phenomenon as in the bulk material i.e due to
interacting localized clusters. Though we are talking about the
nanosystem, however the particles size is big enough to
accommodate lots of atoms. The $T_g$ value is very close in both
N6 and N25, the $T_g$ values do not drastically change with the
particle size but depends on the concentration, is well studied by
Feltin et al .\cite{Feltin}

The field dependance of magnetization has been investigated at 2 K
and 300 K for all the samples and shown in figure \ref{fig:6}. In
case of pure Ni in all the temperature values the M-H curve
exhibit  S  type shape with reasonable coercive field ($H_c$) and
remnant magnetization ($M_r$). This is a characteristic feature of
a spin glass system. The coercivity in the M-H loop is the
indicative of existence of ferromagnetism in the system. The
spontaneous magnetization  value for sample N6 is greater than the
sample N 25 throughout  all the measured temperatures due to large
size of the particle. In case of Ni$_{1-x}$Rh$_x$ at 2 K  all the
samples show coercivity and remnant magnetization however all the
samples do not saturate up to the highest measuring field. The
presence of hysteresis in the of M-H loop at low temperature is
the indicative of ferromagnetic nature of the system. At 300 K,
for x  = 0.27 both the samples show remanent and coercivity
because the samples are ferromagnetic in nature whose $T_C$ lies
above from the room temperature. But for x = 0.64 both the set of
samples show zero remanent and coercivity confirming the samples
are  SPM.  In case of x = 0.64 N6,  at 300 K the sample is SPM
which confirmed from the ZFC-FC with the presence of a blocking
temperature. However for N 25, which is
 paramagnetic as observed from ZFC - FC due to presence of clusters of Ni, we get superparamagnetic nature
 in the M-H. We observe that the  samples don't saturate even at low temp and in  high field and a hysteresis is appeared at low field and low temperature.
 Both, the appearance of hysteresis loop at low temperatures and low field-regions and  non existence of saturation at low temperature in the high field  are the
 characteristic features of spin-glass (SG) \cite{Liang}, \cite{Du} phase with possibly coexisting ferromagnetic clusters.  This gives the clear evidence of existence
 of Ni clusters in the samples whose interaction gives rise to a spin glass state.

 \section{Conclusion}

 In this paper we reported and analyzed the DC  magnetization study of the Ni$_{1-x}$Rh$_x$  nanoalloys of mean diameters 50 - 60
 prepared by a chemical reflux method with two different reaction times. The concentration of the samples were obtained from EDAX measurement.
 The morphology and the size of the particles were determined from SEM studies. The crystallinity and Phase were checked from XRD,  HRTEM images and SAED patterns.
 The chemical states were confirmed from the XPS studies. From the analysis of temperature dependence of magnetization for x $=$ 0.27, ferromagnetism in both the samples are observed.
For x $=$ 0.64, a spin glass phase appeared in both the samples, exhibit a shoulder like feature in M-T curve. The  origin of
spin glass phase is same as that of bulk, due to interaction between  Ni clusters formed due to atomic clustering. However incase of the nanoalloys the short
range interaction  is absent unlike the bulk.

{\bf ACKNOWLEDGEMENTS}
\vspace{10 mm}

 P. Swain acknowledges the financial support from the Council of Scientific and Industrial Research (CSIR), New Delhi.

\end{document}